%% file: Main.tex
\documentclass[lettersize,journal]{IEEEtran}

\usepackage{amsmath,amsfonts}
\usepackage{algorithmic}
\usepackage{algorithm}
\usepackage{array}
\usepackage{textcomp}
\usepackage{stfloats}
\usepackage{url}
\usepackage{verbatim}
\usepackage{graphicx}
\usepackage{cite}

\usepackage{newfloat}
\usepackage{listings}
\usepackage{amssymb}
\usepackage{amsthm}
\usepackage{mathabx}
\usepackage{multirow}
\usepackage{float}

\usepackage{array} 
\usepackage{tabularx}
\usepackage{tikz}
\usepackage{orcidlink} 
\hypersetup{hidelinks} 
\usepackage{colortbl}
\usepackage{arydshln}
\usepackage{booktabs}

\usepackage{pifont}
\newcommand{\xmark}{\ding{55}} 
\newcommand{\cmark}{\ding{51}} 
\usepackage{subfigure}
\usepackage{adjustbox} 
\usepackage{tcolorbox}
\usepackage{titlesec}
\usepackage{subcaption}
\usepackage{cite}

\titlespacing*{\subsubsection}{0pt}{1.5ex plus .2ex minus .2ex}{1ex plus .2ex}

\ifCLASSINFOpdf
\else
\fi

\begin{document}

\title{LLAMA: Multi-Feedback Smart Contract Fuzzing Framework with LLM-Guided Seed Generation}

\author{
Keke~Gai,~\IEEEmembership{Senior~Member,~IEEE},
Haochen~Liang,
Jing~Yu,~\IEEEmembership{Member,~IEEE},
Liehuang~Zhu,~\IEEEmembership{Senior~Member,~IEEE,}
and~Dusit~Niyato,~\IEEEmembership{Fellow,~IEEE}
\IEEEcompsocitemizethanks{
\IEEEcompsocthanksitem K. Gai, H. Liang, and L. Zhu are with the School of Cyberspace Science and Technology, Beijing Institute of Technology, Beijing 100081, China. (E-mails: \{gaikeke, 3220221492, liehuangz\}@bit.edu.cn).
\IEEEcompsocthanksitem J. Yu is with the School of Information Engineering, Minzu University of China, Beijing, China. (Email: jing.yu@muc.edu.cn).
\IEEEcompsocthanksitem D. Niyato is with the College of Computing and Data Science, Nanyang Technological University, Singapore, email: dniyato@ntu.edu.sg.
\IEEEcompsocthanksitem 
This work is supported by the National Natural Science Foundation of China (Grant No.s U24B20146, 62372044).
\IEEEcompsocthanksitem Corresponding author: Jing Yu (jing.yu@muc.edu.cn). 
}
}

\maketitle

\begin{abstract}
\input{Source/Abstract}

\end{abstract}


\input{Source/Intro}

\input{Source/Related}
\input{Source/Method}
\input{Source/Exper}
\input{Source/Coc}

\bibliographystyle{IEEEtran}
\bibliography{sample-base}

\vfill

\end{document}

%% file: Source/Abstract.tex
Smart contracts play a pivotal role in blockchain ecosystems, and fuzzing remains an important approach to securing smart contracts.
Even though mutation scheduling is a key factor influencing fuzzing effectiveness, existing fuzzers have primarily explored seed scheduling and generation, while mutation scheduling has been rarely addressed by prior work.
In this work, we propose a {\em \underline{L}arge \underline{L}anguage Models} (LLMs)-based Multi-feedb\underline{a}ck S\underline{m}art Contr\underline{a}ct Fuzzing framework (LLAMA) that integrates LLMs, evolutionary mutation strategies, and hybrid testing techniques.
Key components of the proposed LLAMA include:
(i) a hierarchical prompting strategy that guides LLMs to generate semantically valid initial seeds, coupled with a lightweight pre-fuzzing phase to select high-potential inputs;
(ii) a multi-feedback optimization mechanism that simultaneously improves seed generation, seed selection, and mutation scheduling by leveraging runtime coverage and dependency feedback; and
(iii) an evolutionary fuzzing engine that dynamically adjusts mutation operator probabilities based on effectiveness, while incorporating symbolic execution to escape stagnation and uncover deeper vulnerabilities.
Our experiments demonstrate that LLAMA outperforms state-of-the-art fuzzers in both coverage and vulnerability detection.
Specifically, it achieves 91\% instruction coverage and 90\% branch coverage, while detecting 132 out of 148 known vulnerabilities across diverse categories.
These results highlight LLAMA’s effectiveness, adaptability, and practicality in real-world smart contract security testing scenarios.

%% file: Source/Intro.tex
\section{Introduction} \label{sec: intro}
Smart contracts, operating as automated protocols on blockchain, execute predefined tasks without requiring mutual intervention \cite{zheng2020overview}. 
As the range of smart contract applications continues to expand \cite{khan2021blockchain}, security vulnerabilities have emerged from multiple angles, attracting considerable attention from both academic and industrial communities.
Our observations indicate that a fundamental security concern arises from the reliance of smart contract execution on their underlying code. 
These codes are prone to vulnerabilities stemming from design flaws or implementation errors, which, once exploited, can compromise assets \cite{kannengiesser2021challenges}. 
Furthermore, the inherent immutability of smart contracts makes post-deployment modifications impractical, even when vulnerabilities are identified.
The escalating logical complexity of smart contracts during development further amplifies the risk of code errors \cite{chu2023survey}. 
Such errors may manifest as omitted boundary condition checks or inconsistencies in data validation. 
Additionally, smart contracts that depend on external data sources are susceptible to malicious manipulation or attacks, leading to erroneous executions. 
For example, attackers exploited crafted transactions to extract significant amounts of Ether from the DAO \cite{mehar2019understanding}, culminating in a hard fork of the Ethereum blockchain.
Given these security challenges, it is crucial for researchers and practitioners to develop robust methodologies for detecting, mitigating, and preventing vulnerabilities in smart contracts.

Previous studies have explored solutions to detecting potential vulnerabilities within smart contracts \cite{vidal2024vulnerability}. 
As one of representative technical tools, fuzz testing addresses the issue of high false positive rates in static analysis~\cite{li2018fuzzing}.
Main challenges of current fuzz testing methods are summarized as follows. 

The first challenge is low-quality or redundant seed generation that causes obstacles in exploring the deeper logic and potential vulnerabilities of the contract. 
Traditional seed generation methods in smart contract rely on pre-configured rules or random generation schemes, which lacks contextual understanding capabilities to generate high-quality and semantically rich initial seeds.
The second challenge is ineffective mutations or resource waste during testing in feedback-driven fuzz testing methods. 
Even though most feedback-driven tools rely on a single feedback signal to optimize a specific phase of the fuzz testing process, e.g., seed generation or mutation scheduling, these tools can hardly integrate multiple feedback signals across the entire fuzz testing process. 
We observe that recent optimizations mainly focus on a specific part of the process rather than overall collaboration throughout the entire testing flow, such that suboptimal test results are caused. 
Finally, the issue of low efficiency and high resource consumption is a great challenge, especially in hybrid fuzzing. 
Despite some efforts in hybrid fuzzing for smart contract vulnerability detection, existing methods typically combine symbolic execution with traditional mutation fuzzing. 
To be specific, symbolic execution generally needs substantial computational resources and longer execution times, which lower down the overall testing efficiency. 
Selecting optimal paths for deep exploration during symbolic execution is a tough job, generally leading to waste of computational resources.

In order to address the issues above, in this paper we propose a {\em \underline{L}arge \underline{L}anguage Models} (LLMs)-based Multi-feedb\underline{a}ck S\underline{m}art Contr\underline{a}ct Fuzzing framework (LLAMA), which addresses the limitations of existing fuzzing methods, i.e., low-quality seed generation, insufficient feedback integration, and inefficiencies in hybrid fuzzing methods.
Our framework consists of three major modules, namely LLM-based Seed Generation (LSG), Multi-Feedback Optimization Strategy (MOS), and Hybrid Fuzzing Engine (HFE) modules.

To be specific, the LSG module employs a hierarchical prompting strategy to guide LLM in extracting critical paths and semantic information from the smart contract, thereby enabling the generation of structurally and semantically valid initial seeds. 
A lightweight pre-fuzzing mechanism is utilized to evaluate and rank the generated seeds, from which the top-k candidates are selected to initialize the fuzzing campaign. 
The final fuzzing results are incorporated as feedback into the prompting process, forming a closed-loop feedback chain for iterative seed improvement.
Next, the MOS module adopts three types of feedback signals to iteratively enhance fuzzing performance across three dimensions, including seed generation, seed selection, and mutation operator scheduling. 
Dissimilar to previous approaches that typically focus on optimizing a single component, our LLAMA systematically integrates feedback across the entire fuzzing workflow. 
Notably, our scheme is among the first to incorporate mutation operator selection based on dynamic feedback, which adaptively adjusts operator probabilities and improves code coverage. 
Finally, the HFE module incorporates symbolic execution as a complementary mechanism to address path stagnation during fuzzing.
When the fuzzing process converges prematurely, symbolic execution is invoked to generate new inputs, thereby enabling deeper state exploration and improving input diversity.

Main contributions of this work are summarized as follows. 
\begin{enumerate} 
\item 
We have proposed a novel initial seed generation framework that utilizes a hierarchical prompting strategy of LLMs to progressively synthesize structured and semantically valid transaction sequences.
The issue of low-quality of redundant seed inputs have been successfully addressed due to our proposed lightweight pre-fuzzing phase and a feedback-aware prompt injection mechanism.
We form a closed-loop seed optimization process, where real-time coverage and vulnerability feedback are embedded into prompt construction, enabling adaptive and vulnerability-sensitive seed generation.


\item 
We are the first to propose a feedback optimization scheme for the seed mutation process in smart contract fuzzing.
To be specific, we propose a comprehensive multi-feedback optimization algorithm through constructing different feedback functions for each major phase of fuzz testing.
Our scheme overcomes the limitation of existing fuzz testing approaches that rely on a single feedback process.


\item 
This work proposes a lightweight hybrid fuzzer based on the selective symbolic execution to address the issue of efficiency in hybrid fuzzing. 
In our approach, due to selectively enabling symbolic execution, the fuzzer only uses symbolic execution when exploring deeper contract paths, such that a significant computation resources are saved. 
The proposed lightweight framework also reduces fuzzer's costs in CPU and memory usage.

\end{enumerate}

The rest of this paper is organized in the following order.
Section~\ref{sec: related} reviews related work.
Section~\ref{sec: method} provides design goals, threat model, and key concepts. 
Sections~\ref{sec:pm} and \ref{sec: exper} details the proposed approach and exhibits experiments evaluations, respectively. 
Finally, Section \ref{sec: coc} concludes this work.

%% file: Source/Related.tex
\section{Related Work} \label{sec: related}

\begin{figure}[t]
\centering
\includegraphics[width=0.48\textwidth]{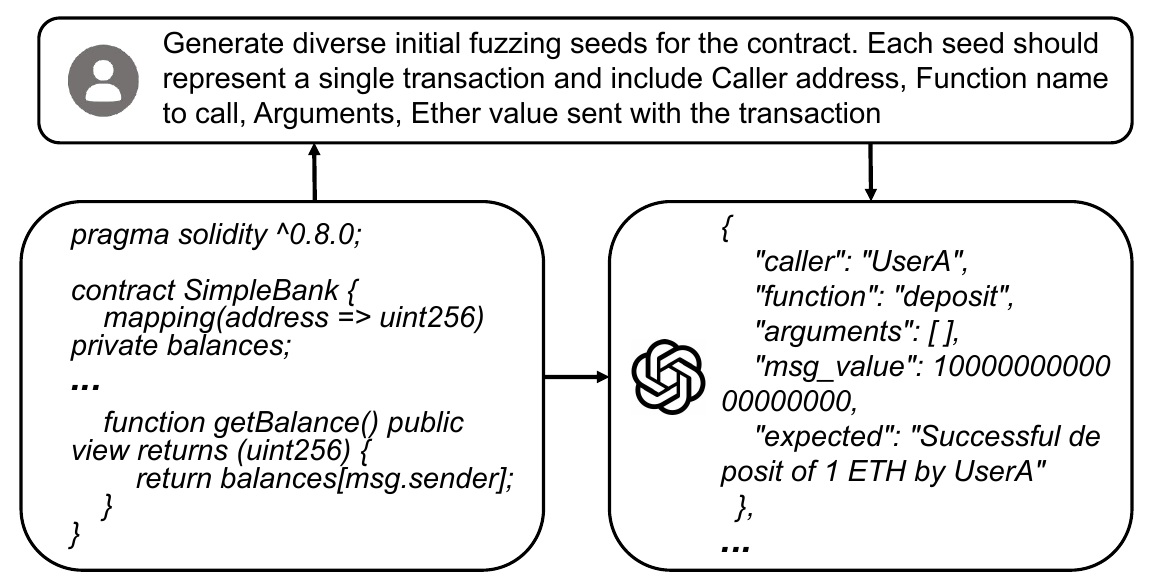} 
\caption{LLM-based seed generation.} 
\vspace{-1.0em}
\label{Fig:LLM-Based fuzzing}
\end{figure}
 
\subsection{LLM-Based Smart Contract Fuzzing}

Compared to existing fuzzers that relied on handcrafted rules or random generation, LLMs was a technical alternative for synthesizing structured and semantically meaningful inputs to enhance code coverage and vulnerability detection performance (refer to Fig.~\ref{Fig:LLM-Based fuzzing}).  
Considering multi-stage prompting for seed generation, prior work \cite{sun2025adversarial} tried a chain-based prompting strategy that divided the seed generation process into multiple stages, e.g., task prompting, context enrichment, review, and optimization. 
An adversarial LLM architecture separated LLM instances and played the roles of generator and evaluator to increase the diversity and effectiveness of generated seeds.

Structured seed optimization and semantic filtering was another direction in the field. 
For example, Codamosa ~\cite{lemieux2023codamosa} utilized LLMs to generate initial inputs and applied multiple rounds of semantic filtering and transformation, for the purpose of ensuring validity and exploitability of test cases.
TrustLLM ~\cite{ma2024combining} guided LLMs with contract summaries and usage contexts to produce interaction sequences that resembled realistic and security-sensitive scenarios.
In addition, other work addressed contract-aware prompting for vulnerability-oriented inputs. 
For instance, CHATAFL ~\cite{meng2024large} and Fuzz4All ~\cite{xia2024fuzz4all}, focused on extracting knowledge from contract documentation, comments, and interaction history, in order to enable LLMs to infer potentially vulnerable execution paths. 
These methods moved beyond ABI-based generation and incorporated richer semantic context into fuzzing input construction.

Even though challenges remain in modeling long-range interactions and generating context-consistent transaction sequences, prior efforts have evidenced that LLMs could improve input quality, test efficiency, and state space exploration.

\begin{figure}[t]
\centering
\includegraphics[width=0.45\textwidth]{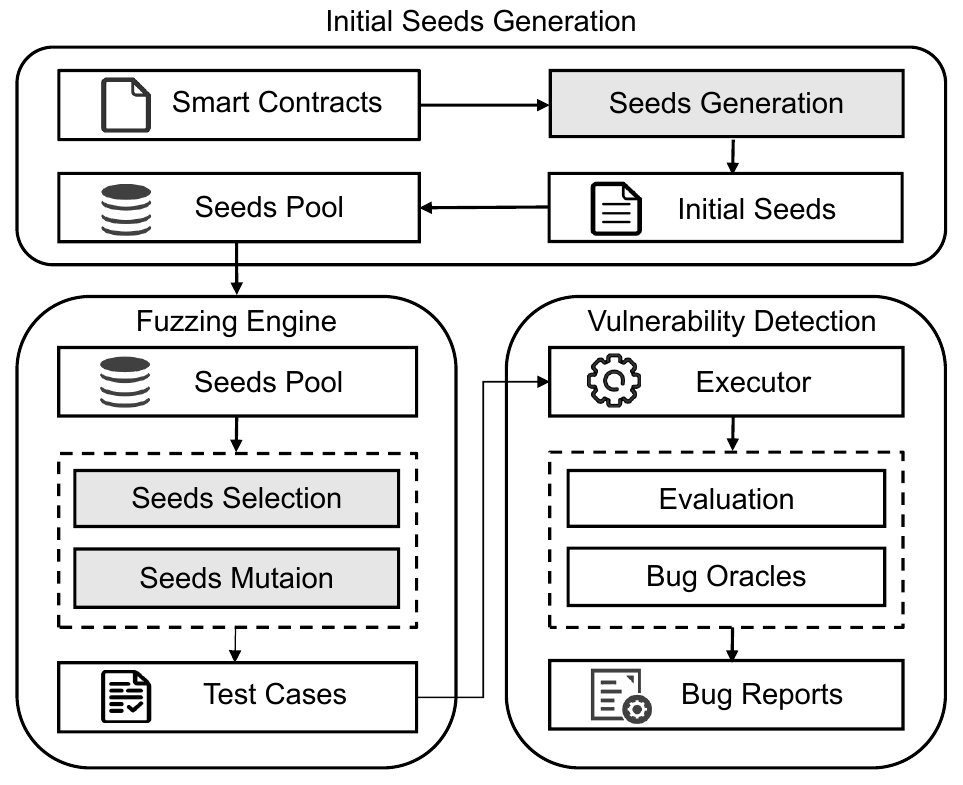} 
\caption{Traditional smart contract fuzzing workflow.} 
\vspace{-1.0em}
\label{Fig:fuzzing}
\end{figure}

\subsection{Feedback-guided Smart Contract Fuzzing}

Feedback-guided smart contract fuzzing is a testing approach for evaluating the security and stability of smart contracts \cite{qian2024mufuzz}. 
A typical such method generates a large volume of random or targeted inputs and leverages runtime feedback to guide input generation and optimization, such as code coverage, execution anomalies, and path exploration data. 
The fuzzer continuously monitors execution traces and refines test cases in terms of feedbacks, with the goal of prioritizing unexplored paths or vulnerability-relevant behaviors \cite{zhu2022fuzzing}. 
As this method deeply examined contract logic and functionality, it had been widely adopted for identifying potential vulnerabilities in smart contracts.

Fig. \ref{Fig:fuzzing} showed traditional smart contract fuzzing that consisted of three key stages, namely, seed generation, selection, and mutation, which determined the fuzzer's performance \cite{wu2024we}. 
In general, seed generations produced high-quality initial seeds capable of triggering deeper execution paths and exposing potential vulnerabilities \cite{herrera2021seed}. 
Seed selections identified valuable seeds from the existing pool for further mutation and testing. 
Seed mutations involved structurally or semantically modifying seeds to generate diverse new inputs \cite{andesta2020testing}.
Previous studies~\cite{nguyen2020sfuzz,choi2021smartian,ji2021increasing} extensively explored feedback mechanisms in the seed selection stage of smart contract fuzzing. 
However, we found that incorporating feedback into the seed generation and mutation stages had been rarely addressed. 
Thus, exploring feedback-driven schemes in above two stages had an urgent demand for further improving the effectiveness of fuzzing.

\subsection{Hybrid Fuzzing for Smart Contract}

Traditional smart contract vulnerability detection methods mainly covered two categories, including the static analysis and fuzz testing. 
To be specific, on the one hand, static analysis identified potential vulnerabilities in smart contracts without executing the code.
Typical techniques included symbolic execution, formal verification, data flow analysis, and machine learning-based methods \cite{tsankov2018securify,tikhomirov2018smartcheck,wang2020contractward}. 
Although static analysis appeared benefits in early-stage vulnerability detections, prior studies also demonstrated that the technique faced issues such as path explosion, high false positive rates, and limited capability in understanding complex contract semantics \cite{mossberg2019manticore,kalra2018zeus,feist2019slither}.
On the other hand, fuzz testing \cite{shou2023ityfuzz} evaluated the security of contracts by executing codes and monitoring runtime behavior. 
Compared with static analysis, this dynamic analysis-based approach had advantages in terms of accuracy in simulating the real execution environment of smart contracts \cite{liu2018reguard,xue2022xfuzz}, and could generally solve the high false positive rate and semantic limitations common in static methods.
However, prior work also tested that fuzz testing faced practical shortcomings, such as low efficiency, blind path exploration, and limited coverage in deep semantic paths \cite{vacca2021systematic}.

Hybrid fuzzing was deemed to be a promising direction in smart contract vulnerability detection that combined strengths of static analysis and fuzz testing. 
Typically, hybrid fuzzing used static insights to guide and optimize the fuzzing process, e.g., control flow, symbolic constraints, and data dependencies.
ILF \cite{he2019learning} formulated the generation of fuzzers as an imitation learning problem, which used symbolic execution to guide input synthesis and enhance early-stage fuzzing performance. 
ConFuzzius \cite{torres2021confuzzius} combined symbolic execution with runtime data dependencies to explore uncovered branches and generate valid transaction sequences. 
Smartian \cite{choi2021smartian} employed static data flow analysis to infer optimal input sequences. 
RLF \cite{su2022effectively} incorporated reinforcement learning to prioritize high-risk execution paths.
Even though hybrid analysis had merits in semantic awareness, coverage, and targeted path exploration, the technique suffered from high-complexity of the system and needed to address complex coordination issues between static and dynamic components. 
The effectiveness of the hybrid framework also heavily depended on the quality of static guidance and run-time feedback; thus, achieving an adaptive balance between the two remains a key research challenge.

%% file: Source/Method.tex
\section{Definitions and the Overview} \label{sec: method}

\subsection{Design Goals} \label{subsec:design_goals}
As a hybrid smart contract fuzzing approach, the major objective of our proposed LLAMA is to address key limitations in existing fuzzers, particularly the inefficiency of uniform mutation strategies and the lack of semantic richness in initial seeds.
Main goals are presented in the following. 

\textbf{High-Quality Seed Generation}:
We aim to generate structurally and semantically valid initial seeds using LLMs, overcoming the limitations of random or rule-based approaches. By incorporating hierarchical prompting and lightweight pre-fuzzing, the framework ensures seeds align with contract logic and prioritize vulnerability-sensitive paths.


\textbf{Adaptive Feedback Integration}:  
We aim to dynamically optimize seed generation, selection, and mutation scheduling through multi-feedback signals (e.g., coverage, dependency analysis). Dissimilar to single-feedback tools, LLAMA integrates runtime insights across the entire fuzzing workflow to maximize exploration efficiency and resource allocation.  

\textbf{Efficient Hybrid Execution}:  
We aim to balance thoroughness and computational cost, i.e., combining evolutionary fuzzing with on-demand symbolic execution. 
The resource needs to be minimized by triggering symbolic analysis only during coverage stagnation.



\subsection{Threat Model} \label{subsec:threat_model}

{\em Adversary's objective}: Attackers aim to exploit potential vulnerability logic within the contract by making specific inputs or interaction flows to carry out illegal activities, including unauthorized access, asset theft, or state manipulation.

{\em Adversary's Knowledge}: Attackers have a complete knowledge of the contract interfaces, including bytecode, ABI specifications, function signatures, and return information, but lacks access to the contract source code, comments, and the developer's business intentions. 
The attack analyzes and reasons about public interfaces and observable execution behaviors without using prior context or private information. 

{\em Adversary's Capability}: The attacker can simulate ordinary users by initiating any legal transactions to the contract and fully control calling functions, input parameters, transaction amounts, calling accounts, and environmental variables (such as timestamps and gas limits), but cannot modify the contract code, disrupt the storage state, or bypass the standard execution process of the {\em Ethereum Virtual Machine} (EVM).

We assume that each component of LLAMA follows a ``semi-honest" policy. 
Specifically, LSG module generates input sequences based on a hierarchical prompt structure, even though it may produce calls with incomplete formats or semantically invalid inputs. 
The fuzz testing engine optimizes path exploration and mutation scheduling in terms of coverage and dependency feedback information without interfering with the contract state.
Symbolic execution in HFE module is only activated when path exploration becomes stagnant, used for solving branch conditions and generating targeted test inputs. 
All testing operations are conducted within an isolated EVM environment to ensure decoupling from the real on-chain execution environment. 

\begin{figure}[t]
\centering
\includegraphics[width=0.48\textwidth]{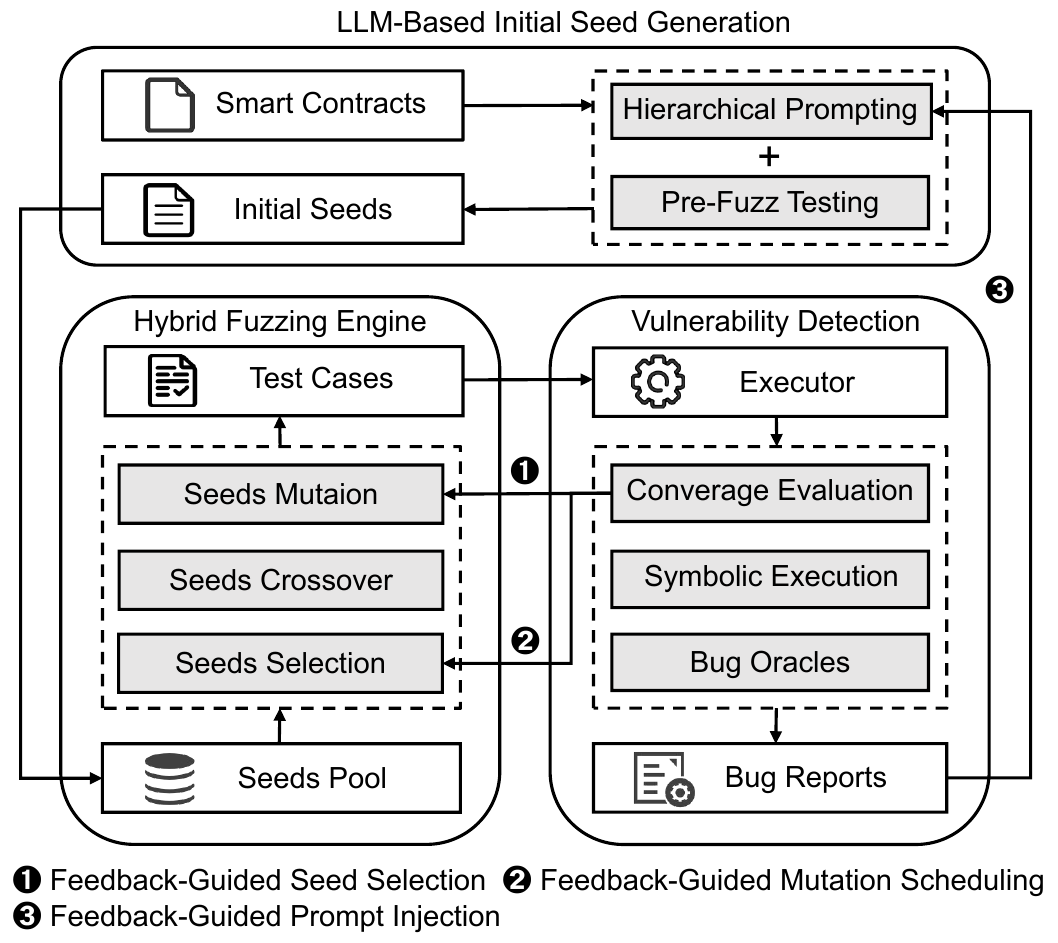} 
\caption{Architecture of the proposed LLAMA.}
\label{fig_archi}
\end{figure}

\subsection{Overview}

The model establishes a multi-layered defense through three core mechanisms.
First, lightweight pre-fuzzing validates format and filters semantics of initial seeds to eliminate potential anomalies from LLM outputs. 
Second, a multi-feedback mechanism enables adaptive evolution of the testing process, reducing path exploration blind spots caused by fixed mutation strategies. 
Finally, selective symbolic execution is to avoid the high resource consumption of traditional hybrid approaches while enabling in-depth analysis of complex path conditions. 
The proposed model allows to effectively counter common attack vectors, such as state dependency spoofing while maintaining testing efficiency.
As shown in Fig.~\ref{fig_archi}, LLAMA consists of three integrated components.

{\bf LLM-Based Initial Seed Generation Module:} 
We adopt a hierarchical prompting strategy to guide large language models in generating structurally valid and semantically meaningful initial test cases. 
A lightweight pre-fuzzing mechanism is further developed to evaluate and filter the generated seeds, ensuring that only high-potential seeds are selected for the main fuzzing phase.
Differing from traditional rule-based or random generated seed inputs, our hierarchical prompting strategy addresses the issue of lacking understanding of smart contracts \cite{li2018fuzzing,sun2025adversarial} and utilizes  the semantic reasoning capabilities of LLMs to achieve the generation of context-aware inputs. 
Inspired by the chained LLM interaction methods, our scheme balances semantic richness and structural validity through phased constraints on the generation process. 
For example, we develop a function abstract layer to extract critical execution paths of contracts to prevent MMLs from generating inputs that deviate from actual logic. 
Our scheme also develops a behavior guidance layer that dynamically adjusts prompts via runtime feedback to address the limitations of traditional methods in modeling long-sequence interactions, e.g., injecting ``generate transactions that modify contract states".

{\bf Multi-Feedback Optimization Strategy Module:} 
To continuously improve fuzzing performance, LLAMA leverages runtime feedback from three dimensions, involving seed generation, seed selection, and mutation operator scheduling. 
These feedback signals are used to iteratively update control parameters across different stages, forming a closed-loop adaptive optimization cycle.
Our investigations find that most existing fuzz testing tools rely on the single feedback, while the complex state dependency of smart contracts requires multi-dimensional signal collaborative optimization, e.g, reflecting data flow validity from RAW dependency. 
Thus, to address the limitation of prior methods, our approach adopts a multi-feedback mechanism using a dynamic-static analysis collaboration in hybrid testing. 
The objective is to dynamically allocate resources and avoid locally optimal solutions through an evolutionary scheme.


{\bf Hybrid Fuzzing Engine:} 
Our fuzzing engine combines genetic algorithms with symbolic execution. 
During multiple testing rounds, high-quality test cases are selected for crossover and mutation. When fuzzing reaches a stagnation point, LLAMA invokes symbolic execution to explore deeper and more complex execution paths.
Differing from existing symbolic execution that lacks efficiency due to path explosion, our proposed LLAMA triggers symbolic execution when the coverage stalls. 
Specifically, the proposed strategy employs a selective symbolic execution that improves the utilization rate of the resources by reducing redundant computations.


By seamlessly integrating LLM-driven input generation, feedback-aware optimization, and evolutionary fuzzing, LLAMA offers a comprehensive solution for smart contract vulnerability detection. It significantly improves both coverage and efficiency, while maintaining high adaptability across a wide range of smart contract scenarios.


\section{Proposed Model}\label{sec:pm}

\subsection{LLM-Based Initial Seed Generation}

\subsubsection{Hierarchical Prompting Strategy}
Traditional fuzzing tools typically rely on symbolic constraints or data dependency analysis to generate a small number of initial transaction sequences as seeds.
This generation scheme exhibits clear limitations in terms of seed diversity and semantic coverage.
To effectively guide the LLM in generating high-quality fuzzing seeds, we propose a five-layer hierarchical prompting strategy that progressively transforms raw contract code into semantically rich and executable test inputs.
\begin{itemize}
    \item 
 \emph{Functional Abstraction:} The LLM first summarizes the purpose and behavior of each function in the contract, capturing side effects, access control modifiers, and key data dependencies. This abstraction provides the semantic foundation for subsequent inference.
    \item 
 \emph{Transaction Sequence Inference:} Based on the functional abstraction, the LLM is prompted to generate valid and diverse sequences of function calls that adhere to intra-contract control and data flow constraints. These sequences are designed to simulate realistic contract usage patterns and explore meaningful execution paths.
    \item 
\emph{Format Verification:} The generated inputs are parsed and validated against the contract's ABI specification to ensure syntactic and semantic correctness, including proper argument types, parameter ordering, and caller context requirements. Ill-formed or infeasible inputs are discarded or corrected.
    \item 
 \emph{Semantic Optimization:} Prompts are further refined to bias the LLM toward producing seeds that are more likely to trigger edge-case logic or vulnerability-relevant paths, such as those involving reentrancy, privilege escalation, or integer overflows.
    \item 
\emph{Behavior-Guided Prompt Injection:} To adaptively align input generation with runtime fuzzing behavior, we introduce a lightweight prompt injection mechanism at the final layer. This component monitors coarse-grained execution metrics and injects a concise, contract-agnostic behavioral hint into the prompt to encourage deeper and more diverse path exploration. 
\end{itemize}

This structured and adaptive prompting pipeline makes LLM synthesize test cases that are not only valid and executable, but also semantically targeted for deep vulnerability exploration. 

\begin{figure}[t]
\centering
\includegraphics[width=0.45\textwidth]{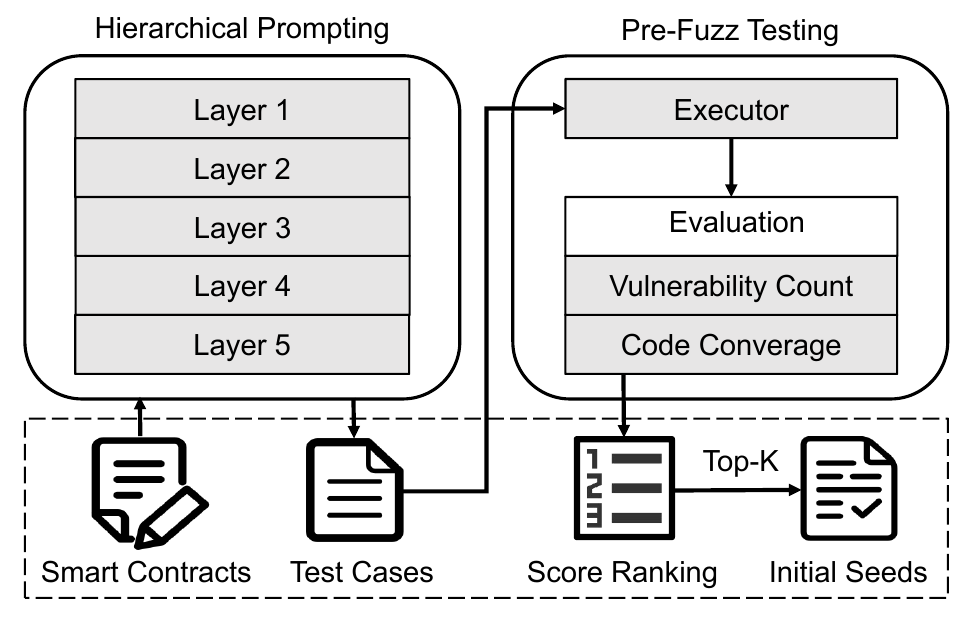} 
\caption{LLM-based seed generation in LLAMA. 
} 
\vspace{-1.0em}
\label{fig3}
\end{figure}

\subsubsection{Pre-Fuzz Testing and Seed Ranking}
To evaluate and filter the seeds generated by LLM, our scheme performs a lightweight pre-fuzzing phase. 
Each seed is executed in an isolated {\em Ethereum Virtual Machine} (EVM) environment to quickly collect basic execution metrics, e.g., instruction-level coverage, call depth, and exception states. 
This process is efficient for ranking large amounts of seeds, as it excludes deep symbolic reasoning or heavy instrumentation.
Based on the collected metrics, we define a scoring function in Eq. (\ref{eq:pre_fuzz_score}), where $\text{Coverage}(s_i)$ represents the number of unique instructions or branches covered by seed $s_i$, $\text{Exception}(s_i)$ is a binary value indicating whether the seed triggered any runtime exception, and $\lambda$ is a small positive constant that slightly boosts seeds which may reveal abnormal behaviors.
\begin{equation} \label{eq:pre_fuzz_score} 
\text{Score}(s_i) = \text{Coverage}(s_i) + \lambda \cdot \text{Exception}(s_i) .
\end{equation}


To select the most promising seeds, we adopt a dynamic Top-$K$ selection strategy. 
Let $N$ denote the total number of seeds generated by the LLM. 
We define $K$ in Eq. (\ref{eq:top_k_selection}), where $\rho \in (0, 1)$ is a proportional factor and $K_{\text{max}}$ is a predefined upper bound on the seed pool size.
\begin{equation} K = \min\left(K_{\text{max}}, \left\lceil \rho \cdot N \right\rceil \right). \label{eq:top_k_selection} \end{equation} 
This scheme ensures that for large $N$, only the top fraction of high-quality seeds are retained, while for small $N$, sufficient diversity is preserved.
The proposed deployment-friendly scheme enables our \textsc{LLAMA} to focus on semantically meaningful inputs while maintaining high throughput and scalability in the seed generation phase.


\subsection{Multi-Feedback Optimization Strategy}

\subsubsection{Feedback-Guided Seed Selection}\label{selection}


    
    
    

\begin{algorithm}[t]
\caption{Feedback-Guided Seed Selection}
\label{alg:SMOSS}
\renewcommand{\algorithmicrequire}{\textbf{Input:}}
\renewcommand{\algorithmicensure}{\textbf{Output:}}
\begin{algorithmic}[1]
\REQUIRE 
    \STATE Seed population $\mathcal{P}$
    \STATE Selection ratio $\gamma \in (0,1)$
    \STATE Coverage analyzer $\mathcal{A}$ 
\ENSURE 
    \STATE Optimized population $\mathcal{P}^*$

\WHILE{not converged}
    \STATE $\mathcal{P}_{\text{new}} \gets \emptyset$ 
    \FOR{each seed $i \in \mathcal{P}$}
        \STATE Execute $i$ and collect metrics:
        \STATE $\quad \Delta_{\text{branch}} \gets \mathcal{A}.\text{NewBranches}(i)$
        \STATE $\quad \Delta_{\text{inst}} \gets \mathcal{A}.\text{NewInstructions}(i)$
        \STATE $\quad \Delta_{\text{RAW}} \gets \mathcal{A}.\text{ValidDependencies}(i)$
        \STATE Compute multi-objective fitness:
        \STATE $\quad \text{fit}(i) \gets \Delta_{\text{branch}} + \Delta_{\text{inst}} + \Delta_{\text{RAW}}$ 
    \ENDFOR
    
    \STATE Rank population:
    \STATE $\quad \mathcal{P} \gets \text{Sort}(\mathcal{P}, \text{fit}(i), \text{descending})$
    \STATE Select top candidates:
    \STATE $\quad k \gets \lceil \gamma |\mathcal{P}| \rceil$
    
    \STATE Generate offspring:
    \STATE $\quad \text{Perform RAW-aware crossover on top-$k$ seeds}$ 
    \STATE $\quad \text{Apply adaptive mutation to offspring}$ 
    
    \STATE Update population:
    \STATE $\quad \text{Replace bottom $(|\mathcal{P}|-k)$ seeds with offspring}$ 
    \IF{No coverage improvement in 5 iterations}
        \STATE Trigger symbolic execution 
    \ENDIF
\ENDWHILE
\RETURN $\mathcal{P}$
\end{algorithmic}
\end{algorithm}

To optimize seed scheduling and enhance evolutionary efficiency, we adopt a feedback-guided seed selection strategy that maintains a pool of candidate seeds (referred to as individuals) in each fuzzing iteration, which are evaluated based on multiple execution-level feedback signals. 
Thus, this strategy enables to consistently prioritize seeds that contribute to increased coverage or generate meaningful behavior.
Moreover, each individual $i \in \mathcal{P}$ is assigned a fitness score $\text{fit}(i)$ that reflects its effectiveness in the fuzzing process, where $\mathcal{P}$ denotes the seed population.
The fitness score is computed by a composite function in the following three dimensions. 
The total fitness score is defined in Eq. (\ref{eq:fitness}).
\begin{equation}
\text{fit}(i) = \Delta_{\text{branch}}(i) + \Delta_{\text{inst}}(i) + \Delta_{\text{RAW}}(i).
\label{eq:fitness}
\end{equation}

\textbf{Branch coverage gain} $(\Delta_{\text{branch}}(i))$: Measures the number of newly explored conditional branches that the seed helps to activate, typically derived from analyzing \texttt{JUMPI} instructions in the EVM trace.

\textbf{Instruction coverage gain} $(\Delta_{\text{inst}}(i))$: Captures the number of low-level EVM instructions triggered by the seed that have not been observed in previous executions.

\textbf{{\em Read-After-Write} (RAW) dependency gain} $(\Delta_{\text{RAW}}(i))$: Reflects the seed’s ability to construct meaningful transaction sequences by establishing read-after-write (RAW) dependencies among storage variables.






Our proposed LLAMA can effectively improve the ability to discover deeper execution paths and complex vulnerabilities in smart contracts, due to integrating runtime feedback into seed selection and utilizing structural (control flow) and semantic (data flow) information.
To be specific, in each iteration of the evolutionary process, our scheme firstly ranks the population in terms of the comprehensive fitness and selects the top $ k = \lceil \gamma |P| \rceil $ elite individuals ($ \gamma $ is the configurable elite retention ratio). 
Next, crossover operations guided by RAW dependencies are performed to recombine high-quality gene segments from elite seeds while ensuring semantic legality of the transaction sequences. 
When a stagnation condition is detected, defined as less than 1\% coverage growth over five consecutive generations, our scheme automatically triggers the symbolic execution module to generate directed test cases that break through path constraints.

\begin{algorithm}[t]
\caption{Adaptive Mutation Scheduling in \textsc{LLAMA} (Proportional Credit Assignment (PCA) strategy)}
\label{alg:AMS}
\renewcommand{\algorithmicrequire}{\textbf{Input:}}
\renewcommand{\algorithmicensure}{\textbf{Output:}}
\begin{algorithmic}[1]
\REQUIRE 
    \STATE Initial seed population $\mathcal{P}$
    \STATE Mutation operator set $\mathcal{M} = \{1,...,m\}$
    \STATE Coverage analyzer $\mathcal{A}$
    \STATE Noise variance $\sigma^2$
    \STATE Population size $\mu$
\ENSURE 
    \STATE Optimized seed population $\mathcal{P}^*$
    \STATE Final operator probability distribution $\mathbf{P}$

\STATE Initialize $\forall j \in \mathcal{M}: fit(j) \gets 0$, $P_0(j) \gets 1/m$
\STATE Analyze initial coverage: $\mathcal{A}.baseline \gets \text{Coverage}(\mathcal{P})$
\WHILE{not TerminationCriterion}
    \STATE $\mathcal{P}_{new} \gets \emptyset$
    \FOR{each parent seed $i \in \mathcal{P}$}
        \STATE Sample a subset of operators $\mathcal{J}_{i'} \subseteq \mathcal{M}$ based on $\mathbf{P}_t$
        \STATE Generate child $i' \gets \text{Mutate}(i, \mathcal{J}_{i'})$
        \STATE Execute $i'$ and collect coverage metrics:
        \STATE $\quad \Delta_{branch} \gets \mathcal{A}.new\_branches(i')$
        \STATE $\quad \Delta_{inst} \gets \mathcal{A}.new\_instructions(i')$
        \STATE Assign credit to each operator:
        \FOR{each $j \in \mathcal{J}_{i'}$}
            \STATE $fit(j) \gets fit(j) + \frac{1}{|\mathcal{J}_{i'}|} \cdot (\Delta_{branch} + \Delta_{inst})$
        \ENDFOR
        \STATE $\mathcal{P}_{new} \gets \mathcal{P}_{new} \cup \{i'\}$
    \ENDFOR

    \STATE Update operator probabilities:
    \FOR{each $j \in \mathcal{M}$}
        \STATE $P_{t+1}(j) \gets \frac{fit(j)}{\sum_k fit(k)} + \mathcal{N}(0,\sigma^2)$
        \STATE $P_{t+1}(j) \gets \max(0.05, \min(0.95, P_{t+1}(j)))$ 
    \ENDFOR

    \STATE Merge populations: $\mathcal{P}_{cand} \gets \mathcal{P} \cup \mathcal{P}_{new}$
    \STATE Select next generation:
    \STATE $\mathcal{P} \gets \text{Top}_\mu(\mathcal{P}_{cand}, \text{score}(i') = \Delta_{branch} + \Delta_{inst})$
    \STATE $\mathbf{P}_t \gets \mathbf{P}_{t+1}$
\ENDWHILE
\RETURN ($\mathcal{P}$, $\mathbf{P}_t$)
\end{algorithmic}
\end{algorithm}

\subsubsection{Feedback-Guided Mutation Scheduling}\label{mutation}

We present an evolutionary operator selection strategy to enhance mutation scheduling under practical fuzzing scenarios.
Rather than statically assigning mutation operators or relying on uniform sampling, \textsc{LLAMA} models the set of mutation operators $\mathcal{M}$ as an evolving population. 
Each operator is treated as an individual entity whose selection probability is dynamically adjusted based on its historical effectiveness.
In each fuzzing iteration, a subset of operators $\mathcal{J}_{i'} \subseteq \mathcal{M}$ is sampled according to a probability distribution $\mathbf{P}_t$ and applied jointly to mutate a parent input. 
The resulting input $i'$ is executed, and two key feedback signals are collected to evaluate the effectiveness of mutation operators as follows.

\textbf{Branch coverage gain} $(\Delta_{\text{branch}}(i'))$: The number of previously unexplored conditional branches (e.g., \texttt{JUMPI} instructions) newly covered by input $i'$. This metric reflects the input’s ability to drive control flow exploration.

\textbf{Instruction coverage gain} $(\Delta_{\text{inst}}(i'))$: The number of EVM instructions triggered by $i'$ that have not been observed in earlier executions. This metric captures the diversity of low-level execution behavior induced by the mutation.



To fairly attribute performance, \textsc{LLAMA} adopts a PCA strategy (refer to Alg. \ref{alg:AMS}), i.e., each participating operator $j \in \mathcal{J}_{i'}$ receives a fraction of the total gain as its fitness increment, denoted by Eq. (\ref{eq:fitness_credit}).
\begin{equation}
\text{fit}(j) \mathrel{+}= \frac{1}{|\mathcal{J}_{i'}|} \cdot \left( \Delta_{\text{branch}}(i') + \Delta_{\text{inst}}(i') \right).
\label{eq:fitness_credit}
\end{equation}
Once fitness scores are updated, a selection mechanism is applied to retain top-performing operators, so that  the selected operators are served for generating the next probability distribution $\mathbf{P}_{t+1}$. 
LLAMA applies a Gaussian perturbation.
\begin{equation}
P_{t+1}(j) = \frac{\text{fit}(j)}{\sum_{k \in \mathcal{M}} \text{fit}(k)} + \mathcal{N}(0, \sigma^2).
\label{eq:probability_update}
\end{equation}
The operator population is thus continuously refined over time, reinforcing useful transformation strategies while enabling the discovery of underutilized or context-sensitive operators.

This evolutionary scheduling mechanism brings two main advantages. First, it allows the fuzzer to prioritize operators that have demonstrated value across diverse execution contexts, reinforcing effective mutation behaviors. Second, by treating mutation operators as evolvable entities, \textsc{LLAMA} achieves a principled balance between exploitation (favoring high-fitness operators) and exploration (retaining operator diversity). Over successive iterations, this strategy drives the fuzzing process toward deeper, more diverse, and vulnerability-relevant paths in the contract's state space.

\begin{figure}[t]
\centering
\includegraphics[width=0.48\textwidth]{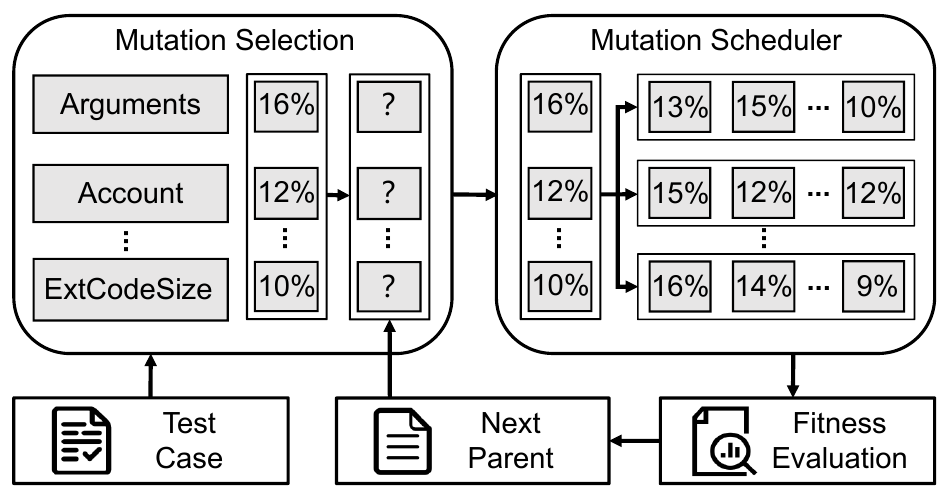}
\caption{Illustration of the evolutionary scheduling process for mutation operator selection in \textsc{LLAMA}.}
\vspace{-1.0em}
\label{fig4}
\end{figure}

\subsubsection{Feedback-Guided Prompt Injection}\label{sec:feedbackinjection}

To finalize the hierarchical prompting strategy in LLAMA, we propose a unified mechanism that integrates runtime feedback with high-level behavioral guidance. This component forms the fifth layer of the prompting framework and establishes a lightweight, contract-agnostic feedback loop between the fuzzing engine and the LLM-based input generator.

Existing prompt adaptation methods generally depend on contract-specific knowledge, e.g., uncovered functions or ABI-based call traces, such that generalizing heterogeneous contract implementation faces a challenge. 
To address this limitation, we abstract the fuzzing feedback into coarse-grained behavioral indicators that remain consistent across different contract structures. 
LLAMA monitors various runtime metrics, covering average call depth, frequency of external calls, Ether transfer occurrences, and whether state-modifying instructions (e.g., \texttt{SSTORE}) are executed.

At the end of each fuzzing iteration, our scheme evaluates behavioral metrics to determine whether the current generation stage exhibits shallow or semantically restricted behavior, e.g., frequent read-only calls, lack of persistent state updates, or missing inter-contract interactions. 
When such a pattern is detected, our scheme selects a natural language behavioral prompt from a predefined pool and appends it to the base prompt. 
These prompts act as soft constraints, guiding LLM to generate inputs that are more semantically rich or more promising to expose vulnerabilities.

The injected feedback remains concise and does not influence the structural consistency established by the earlier prompt layer, e.g., ``Generate a transaction that modifies contract state".
Through this behavior-guided injection strategy, LLAMA forms a closed-loop adaptive mechanism that gradually refines prompt semantics based on observed runtime patterns, which avoid causing significant prompt complexity or contract-specific dependencies.



\begin{algorithm}[t]
\caption{Behavior-Guided Prompt Injection}
\label{alg:behavior_feedback}
\renewcommand{\algorithmicrequire}{\textbf{Input:}}
\renewcommand{\algorithmicensure}{\textbf{Output:}}
\begin{algorithmic}[1]
\REQUIRE 
    Fuzzing behavior metrics $M_t$ from iteration $t$
\ENSURE 
    Behavior-guided prompt hint $h_t$

\STATE // Step 1: Analyze runtime behavior indicators
\STATE Extract:
\STATE \quad Call depth $d$, external call ratio $r$, Ether transfer count $e$, state-write flag $s$ from $M_t$

\STATE // Step 2: Identify shallow or limited behaviors
\IF{$d$ is shallow \textbf{or} $s = 0$ \textbf{or} $r$ is near zero}
    \STATE Select a natural language prompt hint $h_t$ from a predefined hint pool
\ELSE
    \STATE $h_t \gets$ NULL
\ENDIF

\RETURN $h_t$
\end{algorithmic}
\end{algorithm}

\subsection{Hybrid Fuzzing Engine}

\subsubsection{Genetic Algorithm-Based Fuzzing Engine}

The fuzzing engine integrates a genetic algorithm with the goal of generating high-quality transaction sequences, iteratively evolving test inputs through structured selection, crossover, and mutation operations.
In each generation, LLAMA maintains a dynamic population of seeds, in which each individual represents a sequence of transactions interacting with the target smart contract. 
The engine adopts a fitness-based ranking selection strategy to prioritize individuals' performance in code coverage, state diversity, or triggering runtime anomalies. 

Moreover, during the crossover phase, our scheme employs a dependency-aware crossover mechanism, which enforces a RAW constraint when combining two individuals to ensure that meaningful state dependencies exist between transactions. 
Only transaction sequences with cross-function data dependencies are allowed to recombine. This dependency-driven recombination preserves the semantic coherence of the generated sequences and increases the likelihood of triggering complex contract behaviors.

In the mutation phase, LLAMA defines ten mutation operators to perturb function arguments, transaction values, and environmental variables. These mutations are governed by the evolutionary mutation algorithm described in Section~\ref{mutation}, which enhances the reuse of high-value paths while maintaining input diversity. This mechanism facilitates the discovery of rare behaviors and deep vulnerabilities.

LLAMA enables efficient evolution of transaction sequences by integrating dependency-constrained crossover and feedback-aware strategies for selection and mutation.
The engine reinforces existing high-value paths while continuously exploring unexplored regions of the state space to achieve a higher-level semantic coverage and vulnerability detection.

\begin{table*}[t]
\centering
\caption{Mutation Operators and Bug Oracles Supported by LLAMA}
\begin{tabular}{>{\centering\arraybackslash}m{1.5cm}m{6.5cm}|>{\centering\arraybackslash}m{2.2cm}m{5.8cm}} 
    \hline
    \textbf{Mutation Operators} & \textbf{Description} &\textbf{Bug Oracles} & \textbf{Description} \\ \hline
    Arguments & Modifies the parameter values passed during contract invocations to explore distinct logic paths, especially in conditional branches or access control. & Assertion Failure (AF) & Triggered when a failed assertion leads to state changes or fund transfers.\\ \hline
    Account & Mutates the account address involved in the transaction to explore permission-related issues (e.g., administrator vs. regular user). & Block Dependency (BD) & Occurs when contract logic depends on miner-controlled block data like timestamps or block hashes.\\ \hline
    Transaction Amount & Alters the transaction amount (e.g., Ether transferred) to detect potential issues in fund management. & Integer Overflow (IO) & Detected when arithmetic overflows affect contract state or trigger fund transfers.\\ \hline
    Gas Limit & Changes the gas limit to identify gas exhaustion or improper gas management issues during contract execution. & Leaking Ether (LE) & Identified when Ether is sent to untrusted addresses without proper verification.\\ \hline
    Timestamp & Modifies the block timestamp to assess time-dependent logic and miner manipulation of block time. & Freezing Ether (FE) & Happens when the contract can receive Ether but lacks the ability to send it out. \\ \hline
    Block Number & Alters the block number to uncover logic defects related to block height in time-controlled functions. & Reentrancy (RE) & Flagged when repeated external calls allow unintended access to contract state.\\ \hline
    Balance & Modifies the contract or account balance, such as sender and recipient balances, to assess fund management logic. & Transaction Order Dependency (TD) & Detected when contract behavior depends on the order of incoming transactions.\\ \hline
    Call Return Value & Mutates the return values from external contract calls to verify error handling of external interactions. & Unhandled Exception (UE) & Occurs when call failures are not properly checked, leading to unintended execution paths.\\ \hline
    Return Data Size & Modifies the size of return data from external calls to check for handling of varying return data lengths. & Unprotected Selfdestruct (US) & Reported when anyone can trigger selfdestruct without proper access control.\\ \hline
    ExtCodeSize & Changes the size of the external code being called to detect vulnerabilities from incorrect assumptions. & Unsafe Delegatecall (UD) & Triggered when DELEGATECALL targets a dynamic or attacker-controlled address.\\ \hline
\end{tabular}
\label{tab3}
\end{table*}

\subsubsection{Hybrid Fuzzing with Symbolic Execution}

In order to address address issues in satisfying complex or deeply nested path conditions, our scheme forms a hybrid fuzzing framework through integrating symbolic execution as a complementary sector. 
Differing from conventional methods that execute symbolic reasoning as an isolated phase, LLAMA adopts a cooperative execution model, in which symbolic execution is dynamically triggered only when the fuzzing process reaches a plateau/stalls state.

Specifically, when an abnormal situation is detected, LLAMA initiates symbolic execution to analyze those branches that are underexplored/unreachable. 
During this process, the system firstly extracts precise path constraints from complex execution traces. 
The constraints are then solved by using a constraint solver to generate concrete inputs that satisfy the identified paths. 
The resulting inputs are subsequently injected into the seed pool as new individuals, participating in the next generation of mutation and crossover.
Note that the symbolically generated inputs are not treated as isolated test cases. 
Instead, the inputs are fully reintegrated into the evolutionary loop, undergoing the same mutation and recombination processes as other seeds. 
Thus, our scheme allows for retaining the high throughput of fuzzing while using the precision of symbolic reasoning only when necessary.




In contrast to traditional hybrid fuzzers that alternate rigidly between fuzzing and symbolic phases, LLAMA supports a collaborative mode, where both techniques operate in a mutually reinforcing manner. This continuous feedback loop leads to enhanced code coverage and significantly improves the capability to detect vulnerabilities.


\subsection{Mutation Operators and Bug Oracles}

Table~\ref{tab3} summarizes the mutation operators supported by \textsc{LLAMA}, along with their corresponding bug oracles and targeted vulnerability types. 
Each operator is designed to perturb different components of a smart contract's execution context to trigger specific classes of vulnerabilities, such as function arguments, environment variables, or transaction metadata. 
We also define a dedicated detector for each type of vulnerability, which identifies corresponding security issues based on specific patterns or abnormal behaviors observed during the fuzzing process \cite{torres2021confuzzius,qian2024mufuzz}. 
The proposed scheme ensures that each class of vulnerability is detected with precision and effectiveness. 

\input{Source/security}

%% file: Source/security.tex
\subsection{Security Analysis}\label{sec:security}

In line with the threat models defined in Section \ref{subsec:threat_model}, we complete a security analysis in following dimensions. 

Considering the scenario in which adversaries do not have source code access (i.e., the adversary can only access the public interface, e.g., bytecode and ABI), our approach addresses the issue of information asymmetry through combining the semantic guidance of the LSG module and the bybrid execution analysis of the HFE module. 
In details, LSG module uses a hierarchical-hint strategy to guide LLM for generating initial seeds that reflect the potential execution logics and transactions sequences inferred from the contract interface and structure. 
HFE module adopts an analysis method that combines dynamic fuzz testing with selective symbolic execution. 
Thus, our approach does not rely on the source code and evaluates contract behavior by observing the actual execution path, state changes, and symbolic constraints.

In addition, the threat model assumes that adversaries can fully control user-configurable parameters and initiate arbitrary legitimate transactions. 
To address adversarial parameter and interaction controls, LLAMA systematically simulates adversarial capabilities through a diverse set of mutation operators defined in the HFE module, which directly manipulate transaction parameters and environmental variables. 
The MOS module dynamically adjusts the testing strategy in terms of run-time feedback.
The feedback includes the fitness score $fit(i)$ that integrates branch coverage gain $\Delta_{branch}$, instruction coverage gain $\Delta_{inst}$, and RAW dependency gain $\Delta_{RAW}$. 
The feedback-guided mutation scheduling mechanism in MOS updates the selection probability of each operator by its fitness increment. 
Thus, this adaptive quantifiable feedback-driven approach enables the testing behavior to respond to the contract's actual state, which is capable for exploring potential consequences of adversarial parameter manipulation.

Finally, we analyze the reliability of vulnerability detection.
LLAMA checks the state space by integrating evolutionary algorithms with symbolic execution in HFE module. 
The evolutionary process uses fitness evaluations given by MOS to select test sequences and generates new test sequences through dependency-aware crossover and adaptive mutation. 
When the evolutionary process stalls in covering specific paths, HFE triggers symbolic execution to resolve path constraints responsible for the stagnation, thereby overcoming exploration bottlenecks. 
Since all tests are conducted in an isolated environment by EVM specifications and LLAMA's modules follow the ``semi-honest" assumption, it ensures that the detected behaviors are aligned with the intrinsic properties of the contract in standard conditions. 
Therefore, security of the proposed approach depends on the correct deterministic operation of the proposed mechanism within the given security and threat assumptions.

%% file: Source/Exper.tex
\section{Experiments Evaluation} \label{sec: exper}

\subsection{Experiment Configuration} \label{subsec: experset}

{\bf Datasets:}
We selected different datasets for implementing experiments, with the statistical details shown in Table \ref{tab_dataset}. 
Dataset 1 was used to measure the branch coverage and code coverage achieved by LLAMA and various baseline fuzzers. 
The dataset contained 17,803 small-scale contracts and 3,344 large-scale contracts.
Dataset 2 was used to evaluate the performance of different vulnerability detection tools in identifying vulnerabilities. 
The dataset contained a total of 136 contracts and 148 annotat  ed vulnerabilities, covering ten types of vulnerabilities. 
Dataset 3 was used to validate the performance of individual components in LLAMA, which consisted of 500 small contracts and 100 large contracts randomly selected from Dataset 1.

\begin{figure*}[t]
    \centering
    \subfigure[\shortstack{Branch Coverage (Small)}]{
        \includegraphics[width=0.245\linewidth]{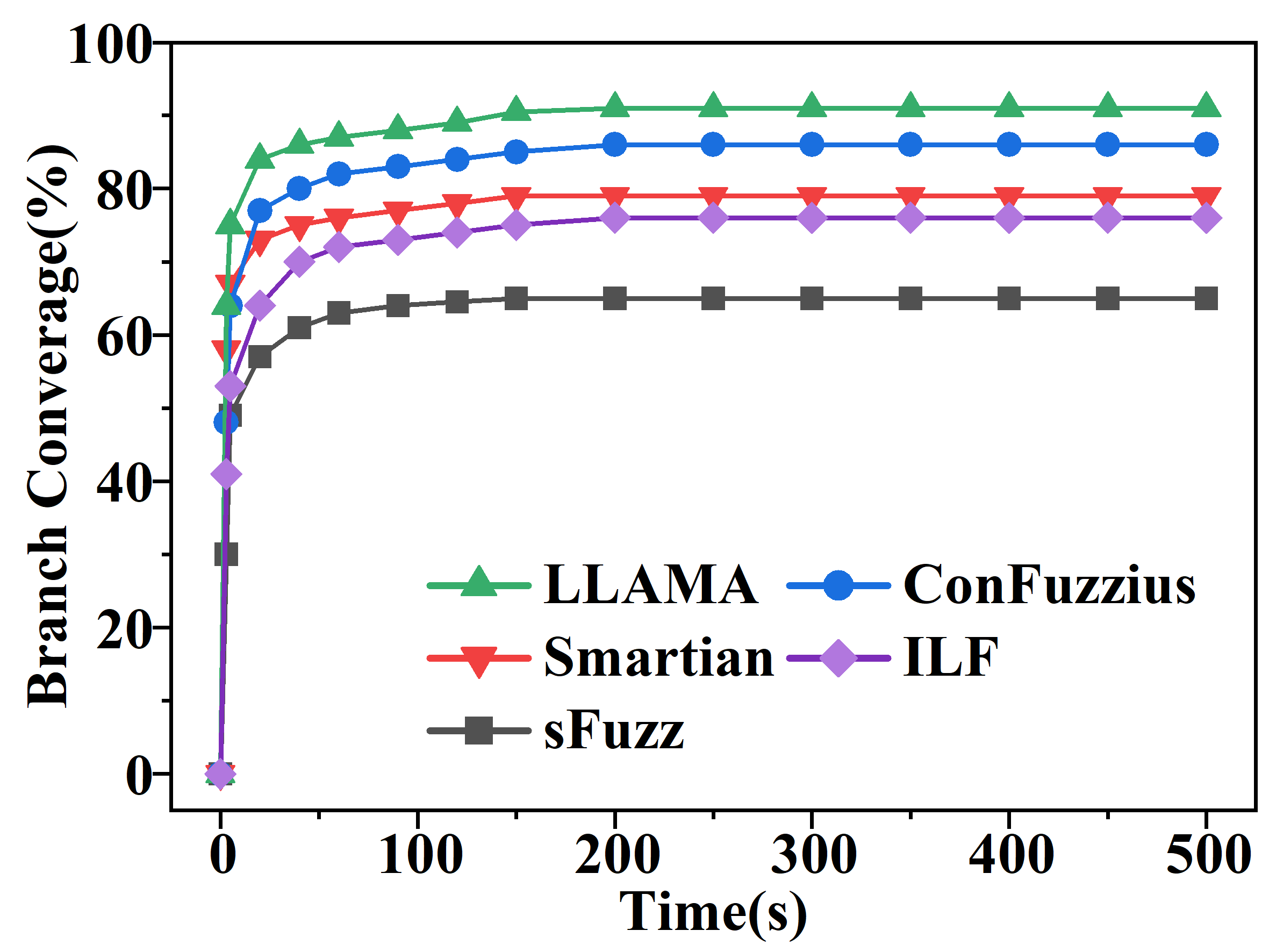}} \hspace{-0.015\linewidth}
    \subfigure[\shortstack{Branch Coverage (Large)}]{
        \includegraphics[width=0.245\linewidth]{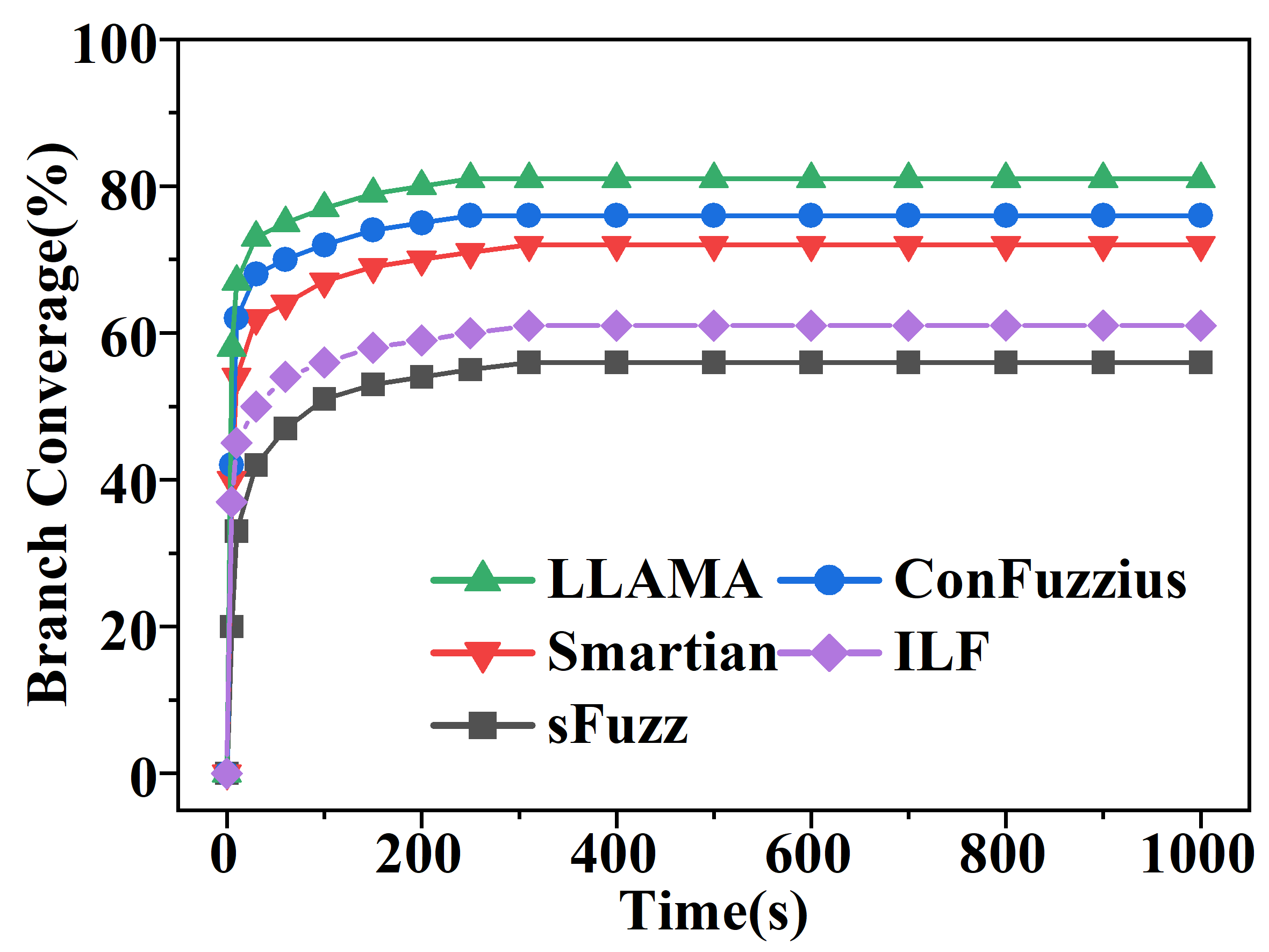}} \hspace{-0.015\linewidth}
    \subfigure[\shortstack{Instruction Coverage (Small)}]{
        \includegraphics[width=0.245\linewidth]{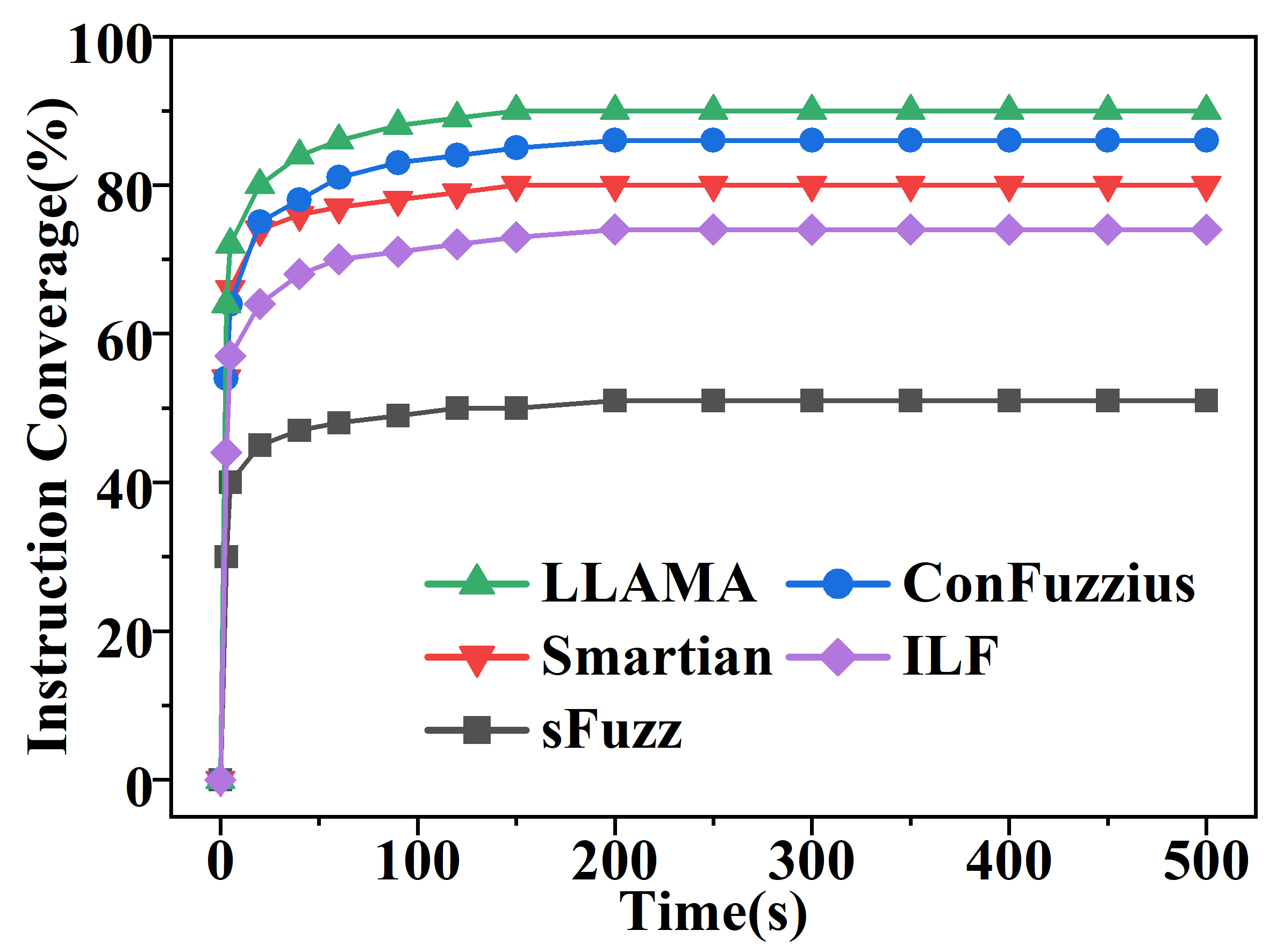}} \hspace{-0.015\linewidth}
    \subfigure[\shortstack{Instruction Coverage (Large)}]{
        \includegraphics[width=0.245\linewidth]{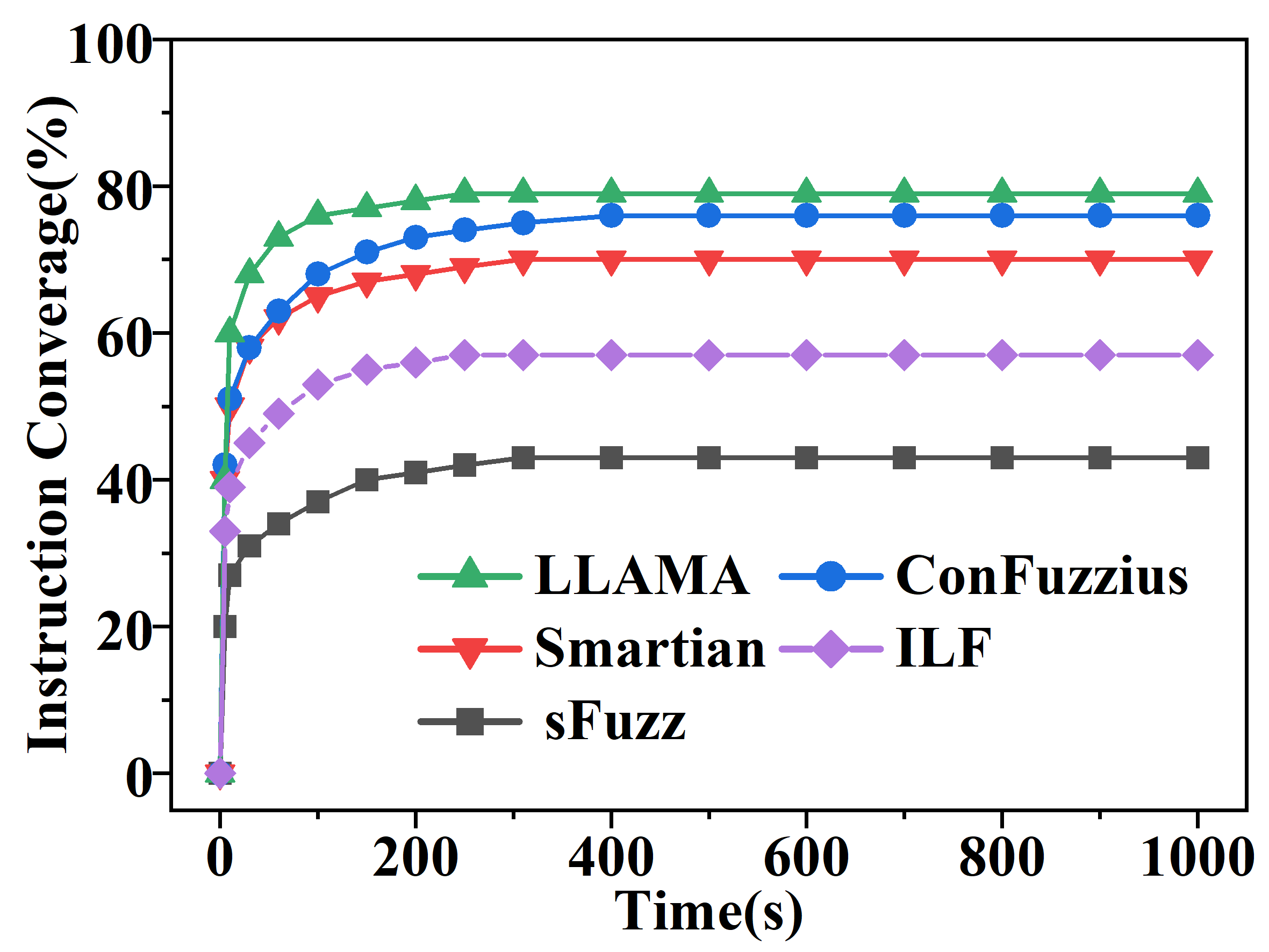}}
    \caption{Branch and instruction coverage comparison on small and large contracts.}
    \label{Fig:coverage}
\end{figure*}

{\bf Baseline:} 
Since LLAMA was a hybrid fuzzing scheme, we compared LLAMA with representative symbolic execution and fuzzing schemes to evaluate whether our approach was more effective than using them independently. 
For symbolic execution schemes, we selected Oyente, Mythril, Osiris, Securify, Slither, and M-Pro, which are highly regarded and widely adopted on GitHub. 
For fuzzing schemes, we selected following baselines.
(1) sFuzz was a representative baseline in smart contract fuzzing. 
(2) ConFuzzius was evaluated as an early hybrid fuzzing scheme that performed the highest code coverage among fuzzers \cite{wu2024we}. 
(3) ILF offered an imitation learning-based fuzzing approach and was evaluated as offering superior performance \cite{wu2024we}.
We compared LLAMA with above fuzzers in terms of instruction coverage, branch coverage, performance, and other relevant metrics.

\begin{table}[t]
    \centering
    \small  
    \caption{Statistics of experimented datasets}
    \begin{tabular}{cp{4.5cm}p{1.5cm}}
        \hline
        \textbf{\#} & \textbf{Source} & \textbf{Scale} \\
        \hline
        Dataset1 & MuFuzz \cite{qian2024mufuzz}, ConFuzzius \cite{torres2021confuzzius}, VeriSmart \cite{so2020verismart}, TMP \cite{zhuang2021smart}  & 21147  \\
        Dataset2 & ConFuzzius \cite{torres2021confuzzius} & 136  \\
        Dataset3 & SmartBugs \cite{durieux2020empirical}, SWC registry \cite{swcregistry} & 155  \\
        Dataset4 & MuFuzz \cite{qian2024mufuzz}, ConFuzzius \cite{torres2021confuzzius}, VeriSmart \cite{so2020verismart}, TMP \cite{zhuang2021smart} & 600 \\
        \hline
    \end{tabular}
   \vspace{-1.0em}
   \label{tab_dataset}
\end{table}

{\bf Experiment Setup:} 
We followed the experiment setup of previous work to achieve a fair evaluation \cite{qian2024mufuzz,torres2021confuzzius}. 
For Dataset 1, we ran each small contract for 10 minutes and each large contract for 20 minutes. 
For Dataset 2, we conducted a 10-minute experiment for each contract. 
For Dataset 3, we also ran each small contract for 10 minutes and each large contract for 20 minutes. 
We ran the experiments on a machine with following configuration, including Ubuntu 20.04 LTS, equipped with two Intel(R) Core(TM) 3090 CPUs at 2.40GHz (32 cores) and 256GB of memory, Solidity version solc-0.4.26 and Z3 solver version 4.8.5, with a timeout of 100ms for each Z3 request.
When code coverage did not increase within 10 generations, LLAMA reinitialized the population.

\subsection{Evaluation on Instruction Coverage and Branch Coverage}

\begin{figure}[t]
\centering
\includegraphics[width=0.9\linewidth]{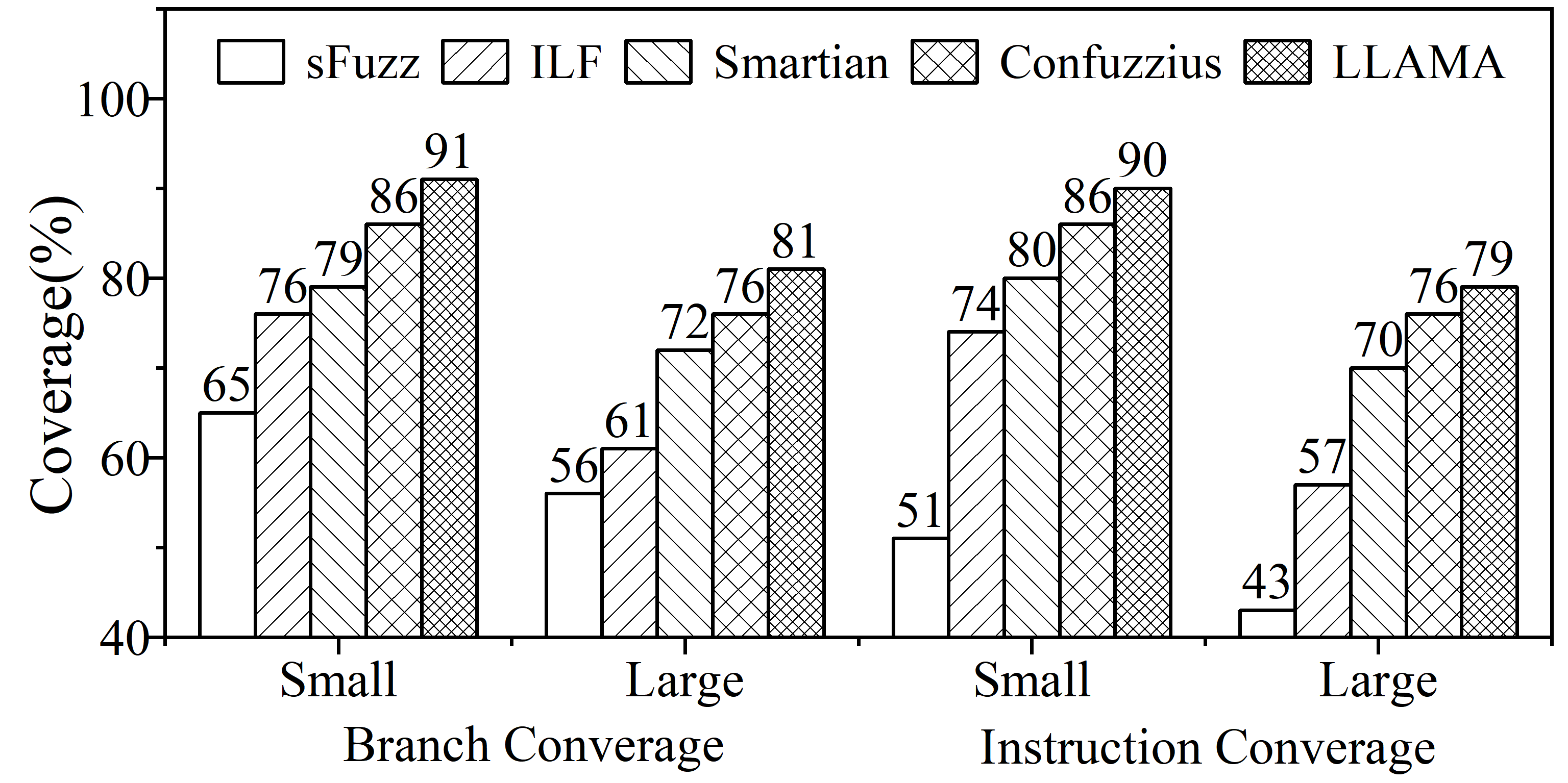}
\caption{Overall coverage comparison} 
\label{Fig:coverageall}
\end{figure}

We first evaluated LLAMA’s coverage metrics, including instruction coverage and branch coverage.
Previous studies have typically chosen only one of these metrics for evaluation, focusing either on instruction coverage or branch coverage. 
However, to provide a more comprehensive assessment of LLAMA’s performance, we evaluated both metrics simultaneously and compared the results with other smart contract fuzzing tools, as shown in Fig. \ref{Fig:coverageall}. 
This dual-metric evaluation not only highlighted LLAMA’s overall effectiveness but also depicted abilities to comprehensively test smart contracts. 
As shown in Fig. \ref{Fig:coverage}, we plotted the coverage-over-time curves to better illustrate the dynamic performance of LLAMA, comparing to other schemes.

The results depicted that LLAMA achieved the highest coverage on both large and small contracts, demonstrating its superior capability in detecting vulnerabilities. 
Specifically, for small contracts, the branch coverage reached 91\%, and the instruction coverage reached 90\%, outperforming all other tools in the comparison.
For large contracts, LLAMA also exhibited strong performance, with branch coverage at 81\% and instruction coverage at 79\%. 
Although the coverage on large contracts was slightly lower than that on small contracts, the difference was less than 11\%, reflecting LLAMA’s consistency in performance across different contract sizes.

From the coverage-over-time curves, LLAMA achieved higher coverage in a shorter period of time compared to other fuzzing tools.
This rapid convergence to high coverage highlighted LLAMA’s efficiency and its ability to detect vulnerabilities early in the fuzzing process. 
The combination of high coverage metrics, minimal performance variance across contract sizes, and faster coverage over time demonstrated that LLAMA not only excelled in absolute performance but also offered a highly practical solution for real-world fuzzing tasks where resources and time were constrained.

\begin{table*}[t]
\centering
\caption{True/False Positives of Each Tool on Different Vulnerability Types}
\begin{tabular}{l|ccccc|ccccc|c}
\hline
\textbf{\multirow{2}{*}{Types}} & \multicolumn{5}{c|}{\textbf{Static Analyzers}} & \multicolumn{5}{c|}{\textbf{Fuzzers}} & \textbf{\multirow{2}{*}{Total}} \\
\cline{2-11}
& Securify & Oyente & Mythril & Slither & M-Pro & Securify & ILF & sFuzz & ConFuzzius & LLAMA &  \\
\hline
AF & n/a & 6 / 6 & 7 / 3 & n/a & 7 / 3 & n/a & n/a & n/a & 10 / 0 & 12 / 0 & 14 \\
BD & n/a & 0 / 0 & 3 / 0 & 3 / 0 & 3 / 0 & n/a & 0 / 0 & 1 / 0 & 7 / 0 & 7 / 0 & 7 \\
LE & n/a & n/a & 4 / 0 & n/a & 4 / 0 & n/a & 4 / 0 & n/a & 4 / 0 & 7 / 0 & 9 \\
FE & n/a & n/a & n/a & n/a & n/a & 3 / 0 & 5 / 0 & 0 / 0 & 5 / 0 & 5 / 0 & 5 \\
IO & n/a & 12 / 4 & 18 / 5 & n/a & 18 / 5 & n/a & n/a & 12 / 0 & 18 / 0 & 19 / 0 & 19 \\
RE & 9 / 0 & 8 / 0 & 10 / 0 & 9 / 0 & 10 / 0 & 4 / 0 & n/a & 7 / 0 & 10 / 0 & 11 / 0 & 11 \\
TD & n/a & 2 / 0 & 0 / 0 & n/a & 0 / 0 & n/a & n/a & n/a & 2 / 0 & 4 / 0 & 4 \\
UD & n/a & n/a & 0 / 0 & 0 / 0 & 0 / 0 & n/a & 1 / 2 & 1 / 2 & 1 / 0 & 1 / 0 & 1 \\
UE & 17 / 0 & n/a & 24 / 0 & 15 / 0 & 24 / 0 & n/a & 10 / 0 & 21 / 0 & 46 / 0 & 63 / 0 & 75 \\
US & n/a & 0 / 0 & 2 / 0 & 3 / 0 & 2 / 0 & n/a & 3 / 0 & n/a & 3 / 0 & 3 / 0 & 3 \\
\hline
\textbf{Total} & 26 / 0 & 28 / 10 & 68 / 8 & 30 / 0 & 68 / 8 & 7 / 0 & 23 / 2 & 42 / 0 & 106 / 0 & 132 / 0 & \textbf{148} \\
\hline
\end{tabular}
\label{tab:final_fuzz_table}
\end{table*}

\begin{figure}[t]
    \centering
    \subfigure[\shortstack{CPU Consumption}]{
        \includegraphics[width=0.48\linewidth]{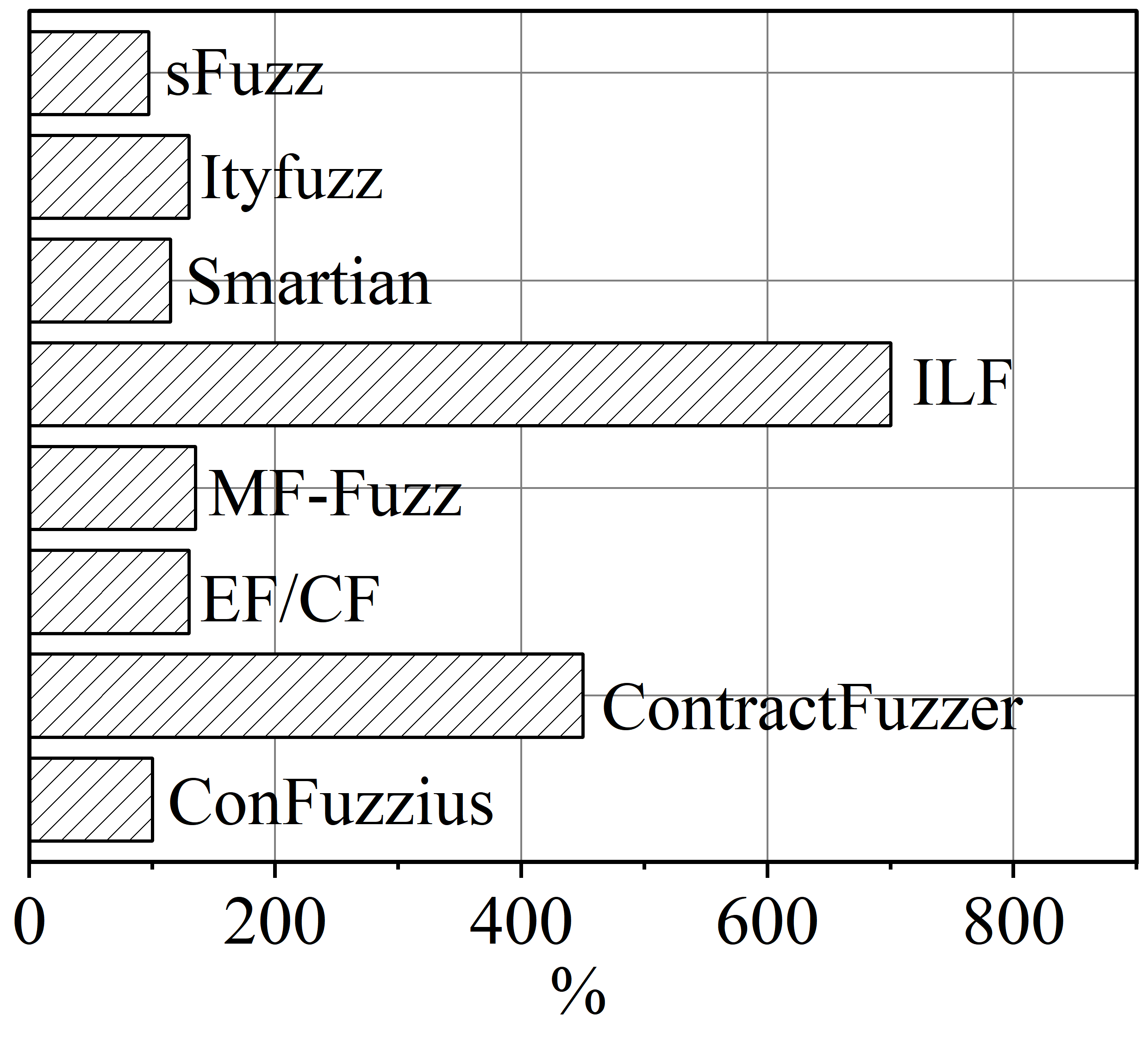}} 
        \hspace{-0.02\linewidth}
    \subfigure[\shortstack{Memory Consumption}]{
        \includegraphics[width=0.48\linewidth]{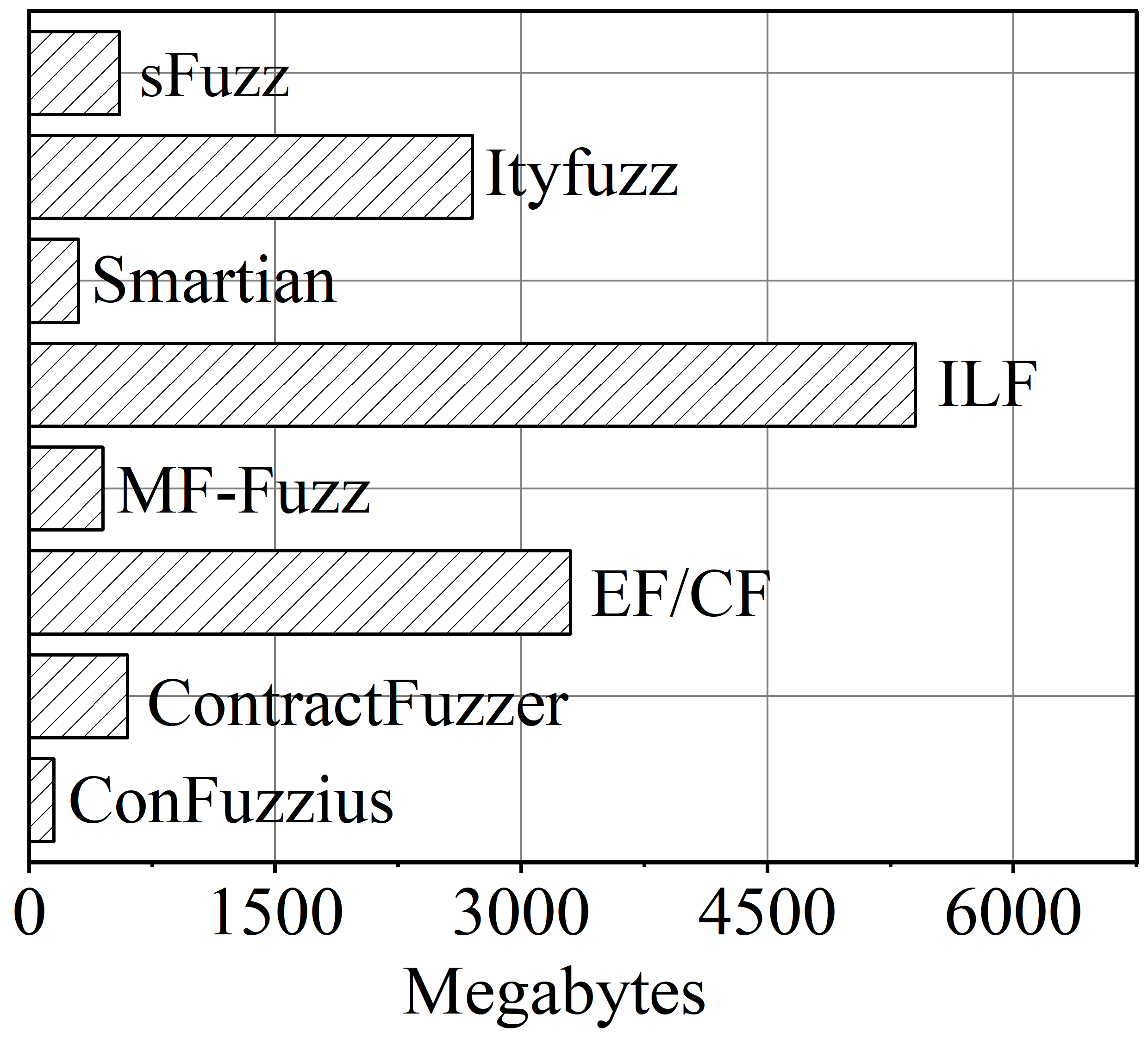}} 
    \caption{Resource consumption comparison.}
    \label{Fig:consumption}
\end{figure}

\subsection{Evaluation on Vulnerability Detection Performance}

We conducted experiments using Dataset 2 to evaluate the vulnerability detection capabilities of different approaches.
Table \ref{tab:final_fuzz_table} summarizes the number of vulnerabilities of each type detected by each tool, along with the reported true positives and false positives.
It was evident that LLAMA detected significantly more vulnerabilities than other tools, demonstrating its superior capability in detecting multiple types of vulnerabilities.
In detail, LLAMA detected 132 out of 148 vulnerabilities, achieving a detection rate of 89\%, which was 18\% higher than the baseline approach, ConFuzzius, which reached 71\%. 
For each type of vulnerability, LLAMA also identified the highest number of vulnerabilities.

LLAMA reported the most true positives and the fewest false positives.
For example, in the case of unhandled exception vulnerabilities, LLAMA detected 63 vulnerabilities, representing a 23\% increase in detection compared to the state-of-the-art ConFuzzius.
Moreover, referring to Tables \ref{tab3} and \ref{tab:final_fuzz_table}, LLAMA achieved further improvements for other types (e.g., AF, IO, RE, TD, BD, UD, FE, US) compared to other approaches.
LLAMA achieved 100\% detection accuracy on datasets for IO, RE, TD, BD, UD, FE, and US types of vulnerabilities.
Overall, the experiment results indicated that LLAMA performed best in detecting a wide range of vulnerabilities.
These findings underscore the importance of LLAMA in enhancing detection capabilities for multiple types of vulnerabilities.

\subsection{Evaluation on CPU and Memory Overhead}
Fig.~\ref{Fig:consumption} showed comparisons on overhead of CPU and memory via different fuzzing tools.
The results showed that ILF incurred the highest CPU and memory overhead due to its reliance on a complex deep learning model, making it less practical in environments with limited hardware resources.
ConFuzzius demonstrated the lowest CPU and memory overhead, benefiting from its lightweight EVM design.
Similar to ConFuzzius, our approach adopted a lightweight EVM within a hybrid fuzzing framework, but it showed a slight additional overhead compared to ConFuzzius.
We found that the slight overhead was due to the integration of LSG and MOS modules.
Our evaluations showed that LLAMA's cost was negligible and within an acceptable limit, significantly lower than other approaches, demonstrating a great balance between performance and resource efficiency.

\begin{figure*}[t]
    \centering
    \subfigure[\shortstack{Branch Coverage (Small)}]{
        \includegraphics[width=0.245\linewidth]{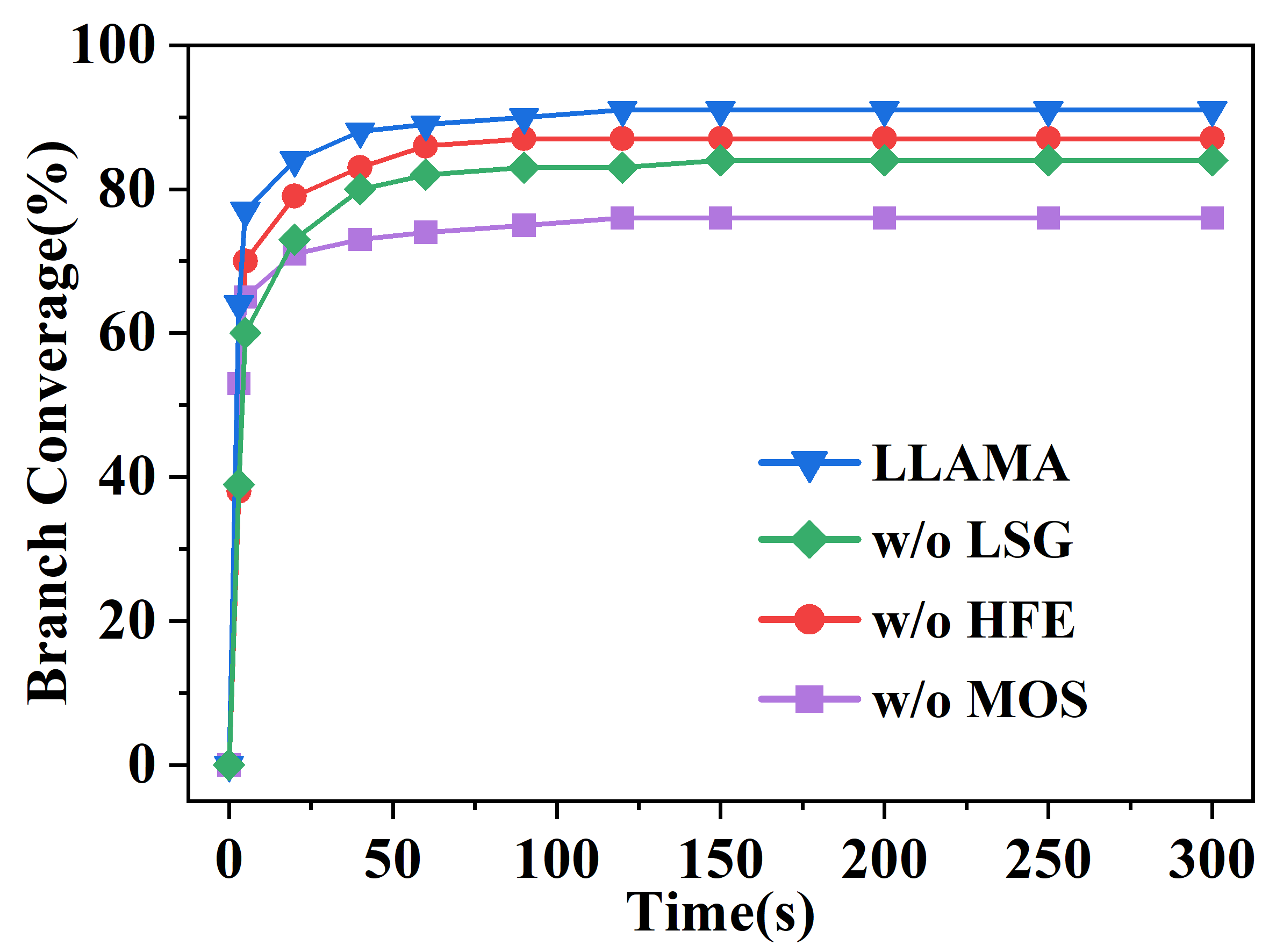}} \hspace{-0.015\linewidth}
    \subfigure[\shortstack{Branch Coverage (Large)}]{
        \includegraphics[width=0.245\linewidth]{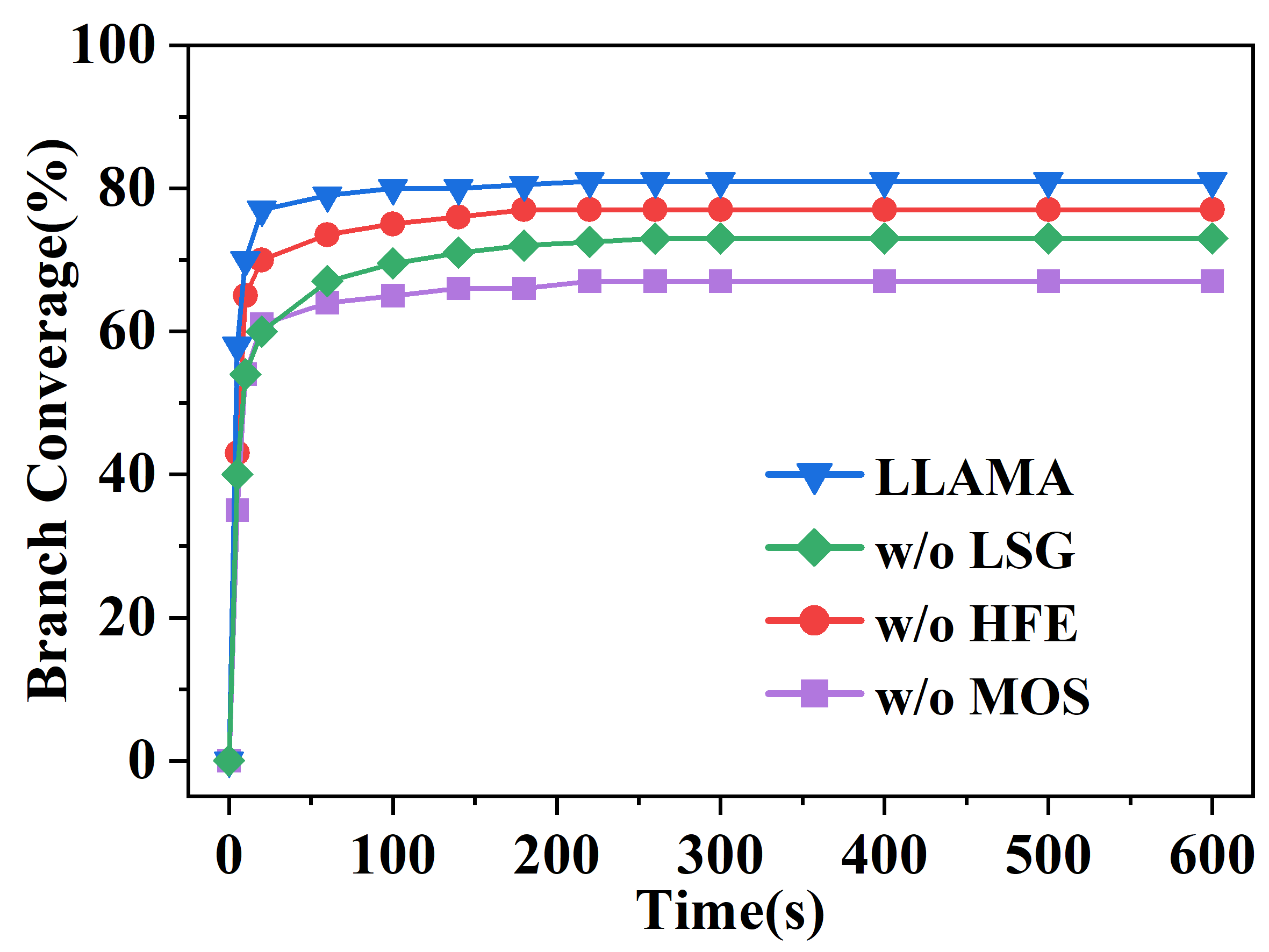}} \hspace{-0.015\linewidth}
    \subfigure[\shortstack{Instruction Coverage (Small)}]{
        \includegraphics[width=0.245\linewidth]{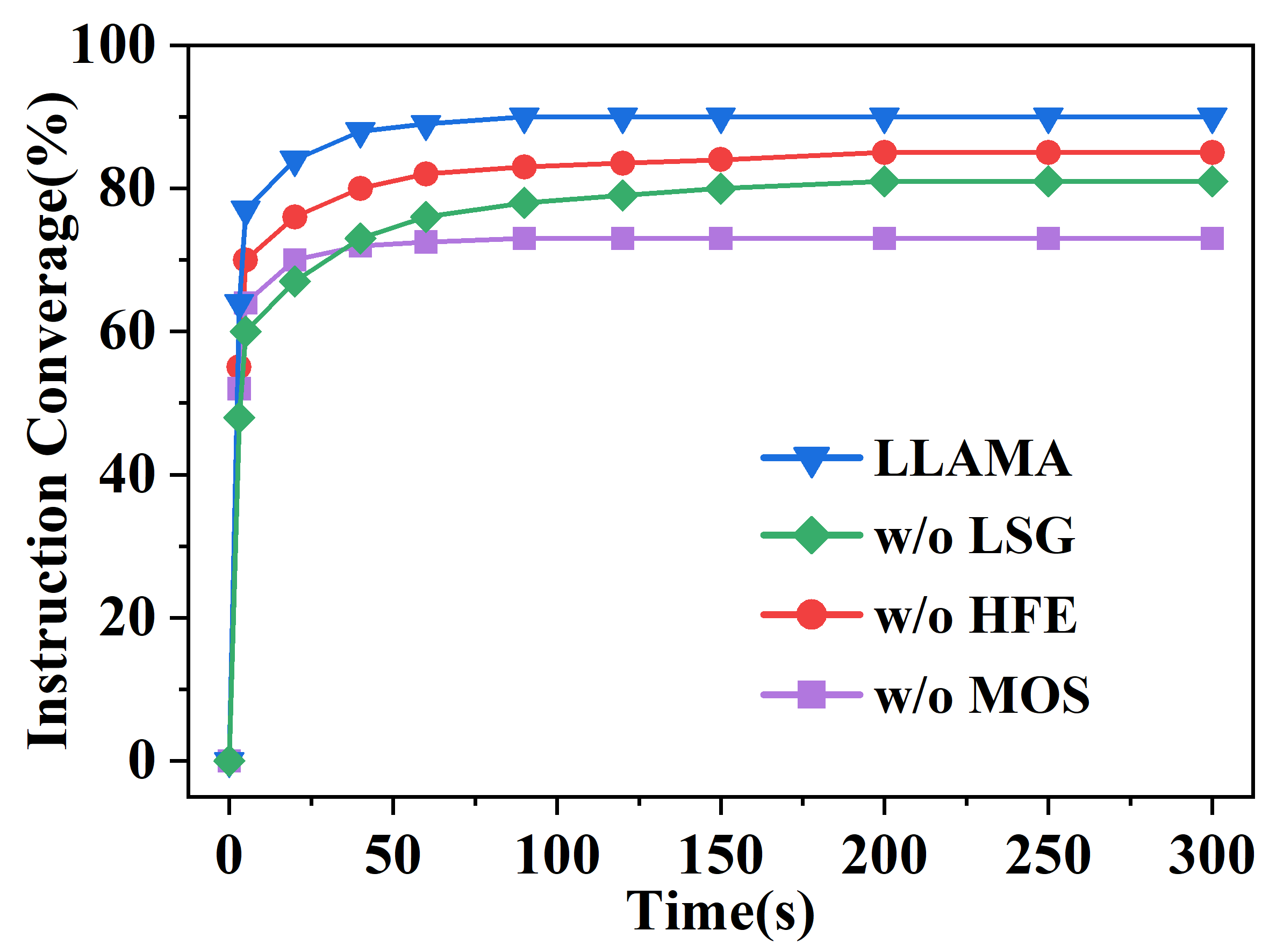}} \hspace{-0.015\linewidth}
    \subfigure[\shortstack{Instruction Coverage (Large)}]{
        \includegraphics[width=0.245\linewidth]{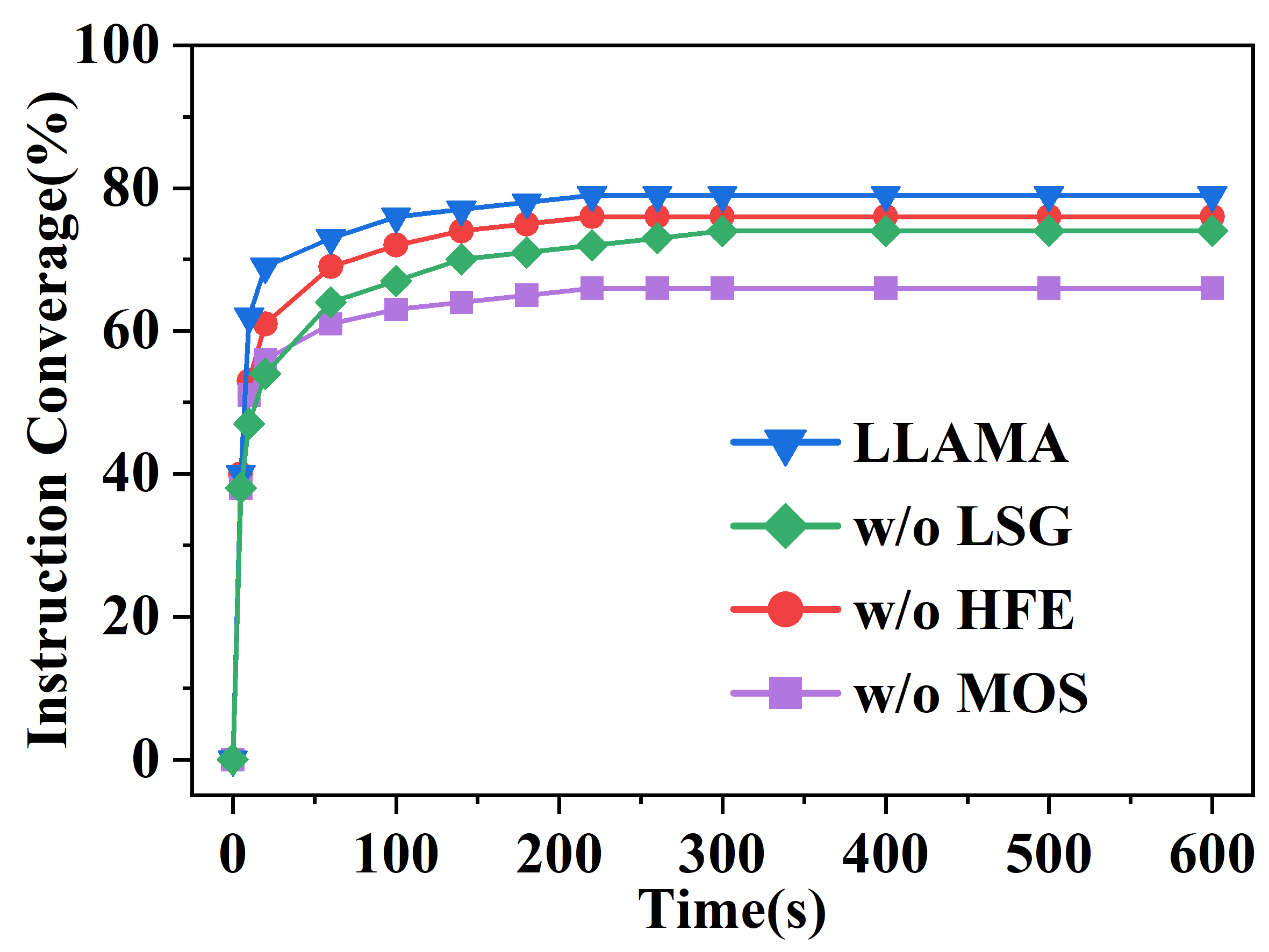}}
    \caption{Branch and instruction coverage comparison on small and large contracts.}
    \label{Fig:ablation}
\end{figure*}

\begin{figure}[t]
\centering
\includegraphics[width=0.9\linewidth]{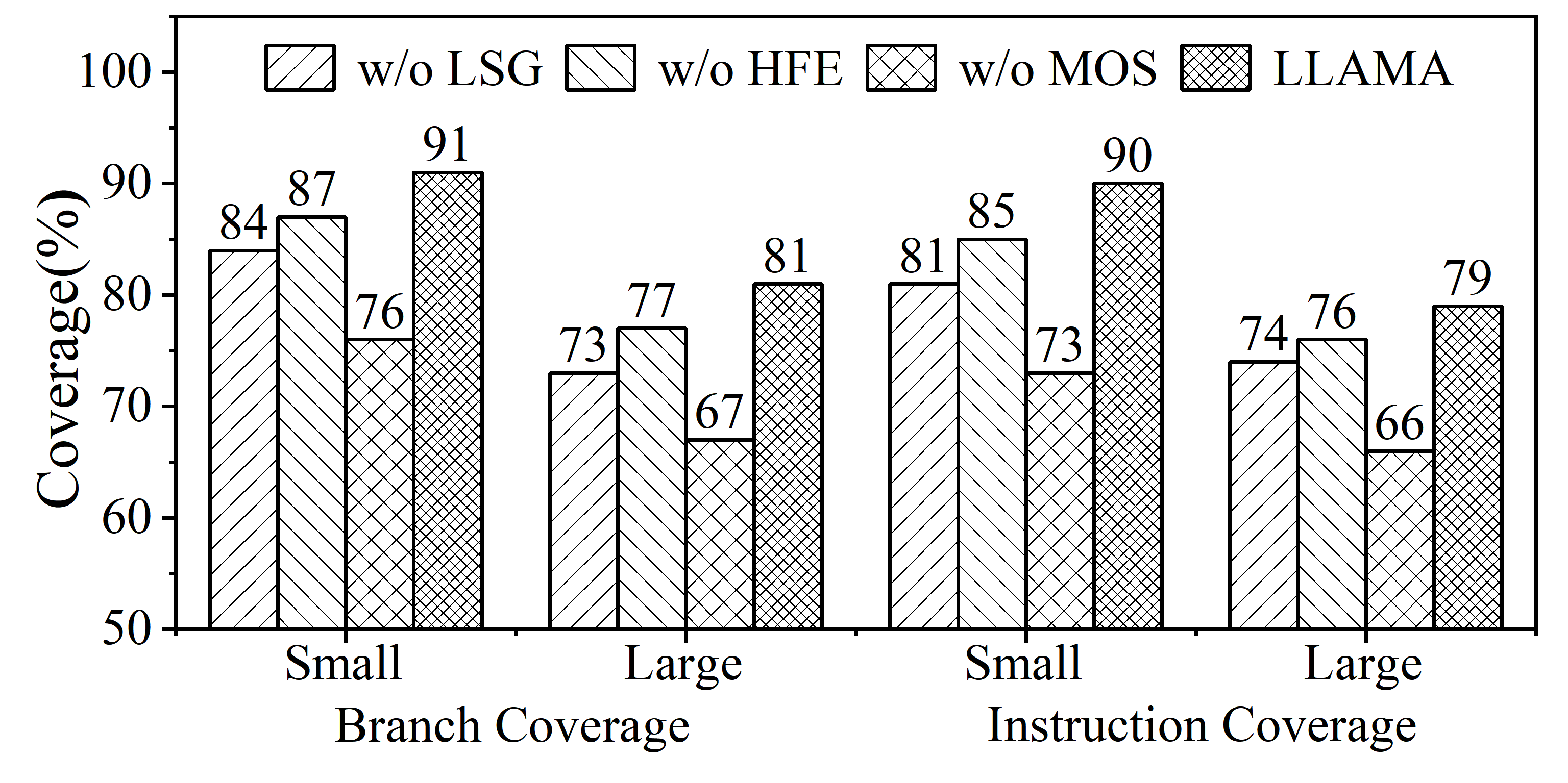}
\caption{Overall ablation study.} 
\label{Fig:ablationall}
\end{figure}

\subsection{Ablation Study}
To evaluate the contribution of each core component in LLAMA, we conducted an ablation study by individually disabling its three key modules, i.e., the LSG, MOS, and HFE modules. 
We evaluated four configurations: LLAMA without the LSG module (w/o LSG), without the HFE module (w/o HFE), without the MOS module (w/o MOS), and the complete LLAMA framework.
The experiments were conducted on two datasets consisting of 500 small-scale and 100 large-scale smart contracts. Branch coverage and instruction coverage were adopted as the evaluation metrics.
Figs.~\ref{Fig:ablationall} and \ref{Fig:ablation} illustrate the overall coverage results and the coverage-over-time curves for the four configurations, respectively.

The results showed that a complete LLAMA significantly outperformed all its variants across both datasets and metrics.
Removing the LSG module led to a noticeable drop in coverage and significantly slowed the coverage growth in the early stages, indicating that high-quality initial seeds generated by the LLM were essential for accelerating early path exploration.
Disabling the HFE module also resulted in reduced coverage performance, suggesting that the HFE module contributes positively to mutation effectiveness.
Removing the MOS module caused the most substantial decline in coverage and led to earlier stagnation in coverage growth, demonstrating that the MOS module played a critical role in maintaining long-term coverage improvement.

Overall, our proposed LLAMA not only achieved superior overall coverage but also demonstrated faster convergence during the fuzzing process.

\begin{figure}[t]
    \centering
    \subfigure[\shortstack{Execution times.}]{
    \label{fig:execution-1}
        \includegraphics[width=0.48\linewidth]{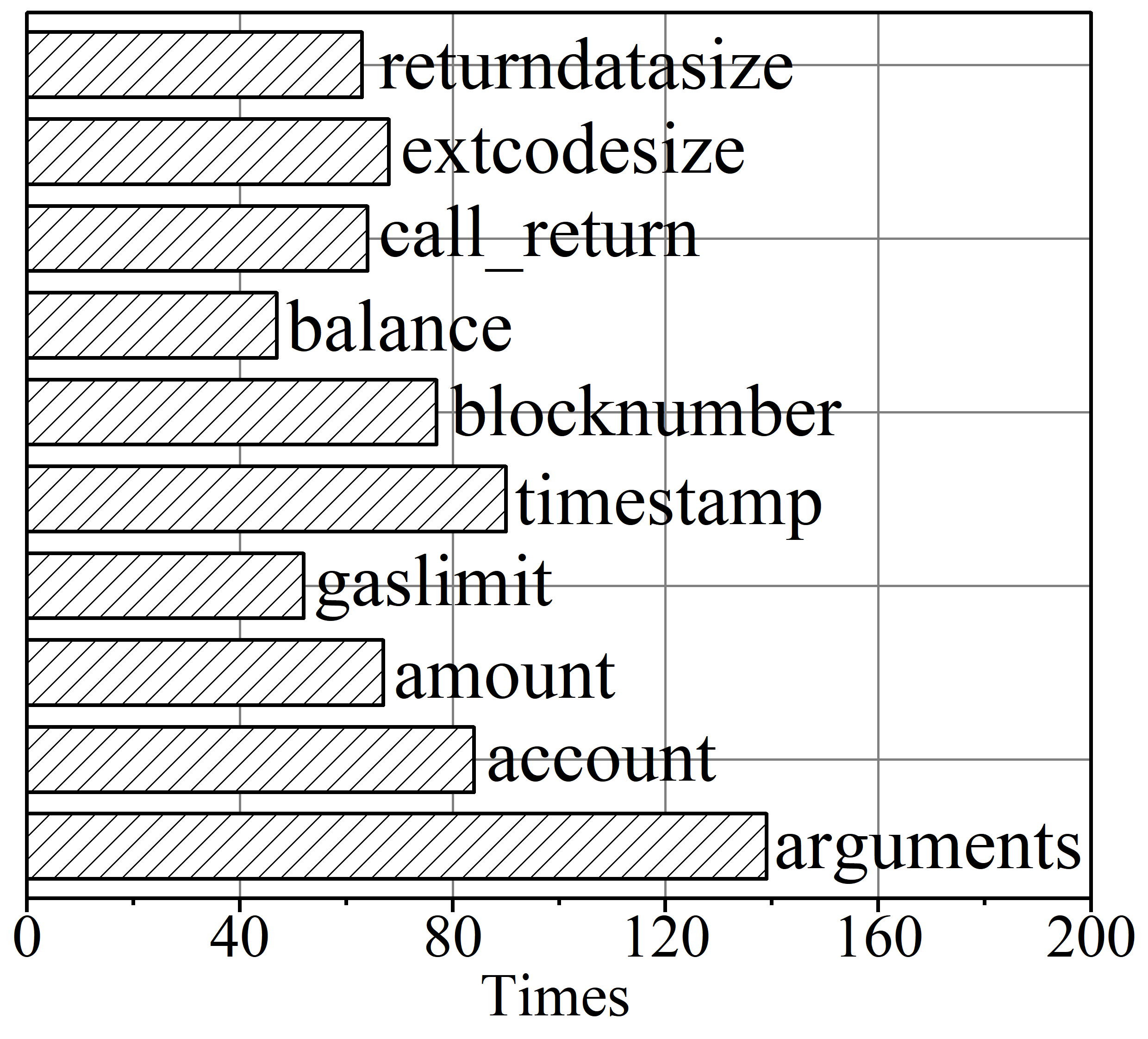}} 
        \hspace{-0.02\linewidth}
    \subfigure[\shortstack{Coverage growth.}]{
    \label{fig:execution-2}
        \includegraphics[width=0.48\linewidth]{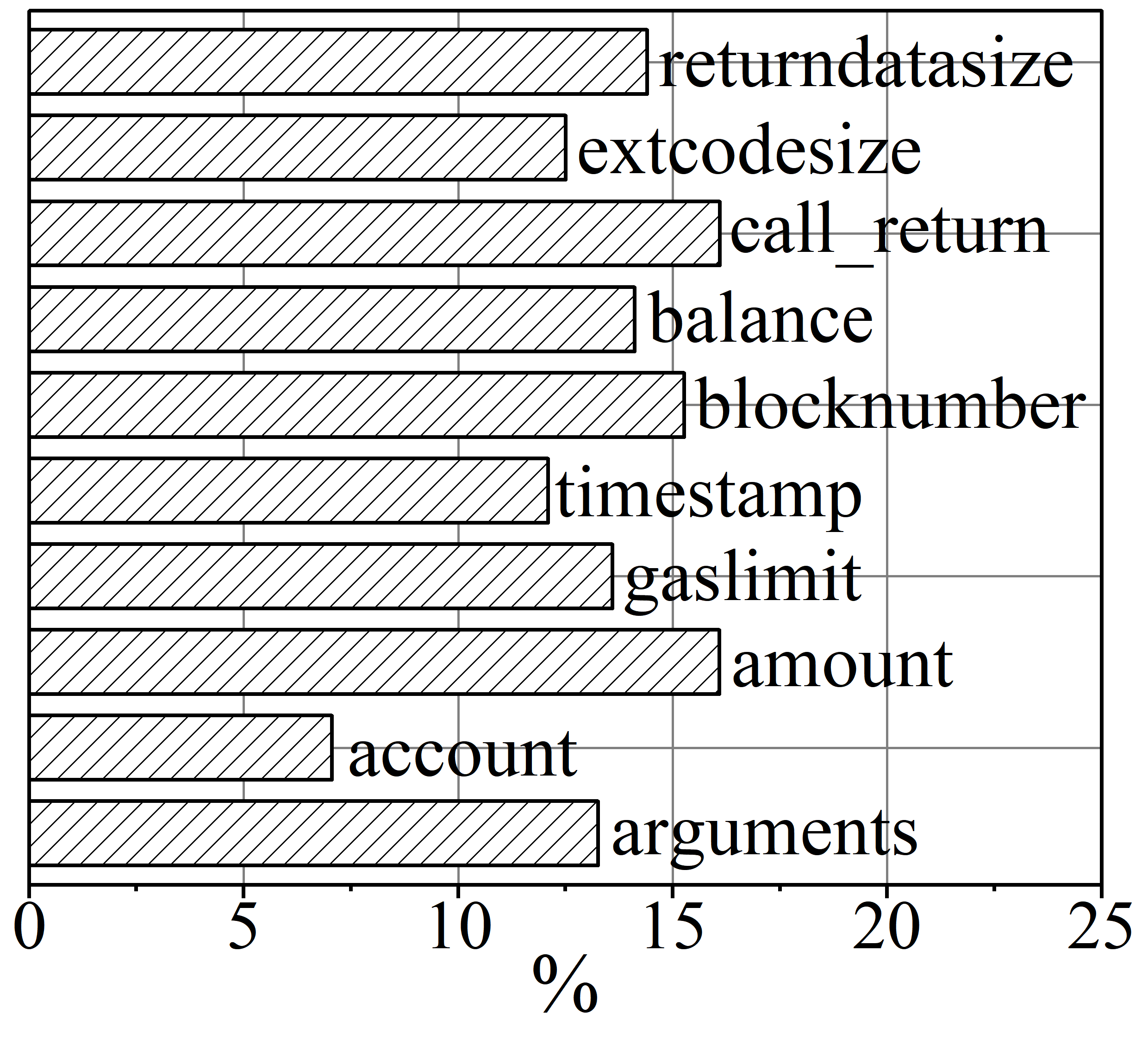}} 
    \caption{Execution times and coverage growth of different mutation operators.}
    \label{Fig:mutaion}
\end{figure}

\begin{table*}[t]
\centering
\caption{Comparison of Existing Tools for Smart Contract Vulnerability Detection}
\begin{tabular}{lcccccccccccc}
\hline
\multirow{2}{*}{\textbf{Tool}} & \multirow{2}{*}{\textbf{Type}} & \multirow{2}{*}{\textbf{Public Tool}} & \multicolumn{10}{c}{\textbf{Vulnerability Type}}\\
\cmidrule(lr){4-13}
 &  &  & \textbf{AF} & \textbf{BD} & \textbf{EL} & \textbf{FE} & \textbf{IO} & \textbf{RE} & \textbf{TD} & \textbf{UD} & \textbf{UE} & \textbf{US}  \\
\hline
Maian \cite{nikolic2018finding}& Static Analyzer & YES & \xmark & \xmark & \xmark & \xmark & \cmark & \xmark & \xmark & \xmark & \xmark & \cmark \\

SmartCheck \cite{tikhomirov2018smartcheck} & Static Analyzer & YES & \xmark & \cmark & \xmark & \cmark & \cmark & \cmark & \xmark & \xmark & \cmark & \xmark \\

Slither \cite{feist2019slither} & Static Analyzer & YES & \xmark & \cmark & \cmark & \cmark & \xmark & \cmark & \cmark & \cmark & \cmark & \cmark \\

Manticore \cite{mossberg2019manticore} & Static Analyzer & YES & \cmark & \cmark & \xmark & \xmark &  \cmark & \cmark & \xmark & \xmark & \cmark & \cmark \\

DefectChecker \cite{chen2021defectchecker} & Static Analyzer & YES & \xmark & \cmark & \xmark & \cmark & \xmark & \cmark & \xmark & \xmark & \cmark & \xmark \\

Mythril \cite{mueller2017mythril} & Symbolic Executor & YES & \cmark & \cmark & \cmark & \xmark & \cmark & \cmark & \cmark & \cmark & \cmark & \cmark \\

M-Pro \cite{mueller2017mythril} & Symbolic Executor & YES & \cmark & \cmark & \cmark & \xmark & \cmark & \cmark & \cmark & \cmark & \cmark & \cmark \\

Osiris \cite{torres2018osiris} & Symbolic Executor & YES & \xmark & \cmark & \xmark & \xmark & \cmark & \cmark & \xmark & \xmark & \xmark & \xmark \\

Oyente \cite{luu2016making} & Symbolic Executor & YES & \cmark & \cmark & \xmark & \xmark & \cmark & \cmark & \cmark & \xmark & \xmark & \cmark \\

teEther \cite{krupp2018teether} & Symbolic Executor & YES & \xmark & \xmark & \xmark & \xmark & \xmark & \xmark & \xmark & \cmark & \xmark & \cmark \\

ContractFuzzer \cite{jiang2018contractfuzzer} & Fuzzer & YES & \xmark & \cmark & \xmark & \cmark & \xmark & \cmark & \xmark & \cmark & \cmark & \xmark \\

ContraMaster \cite{wang2020oracle} & Fuzzer & YES & \xmark & \xmark & \xmark & \xmark & \cmark & \cmark & \xmark & \xmark & \cmark & \xmark \\

Echidna \cite{grieco2020echidna} & Fuzzer & YES & \xmark & \xmark & \xmark & \xmark & \xmark & \xmark & \xmark & \xmark & \cmark & \xmark \\

Reguard \cite{liu2018reguard} & Fuzzer & NO & \xmark & \xmark & \cmark & \xmark & \xmark & \xmark & \xmark & \xmark & \xmark & \xmark \\

Harvey \cite{wustholz2020harvey}& Fuzzer & NO & \cmark & \xmark & \xmark & \xmark & \cmark & \cmark & \xmark & \xmark & \cmark & \xmark \\

sFuzz \cite{nguyen2020sfuzz}& Fuzzer & YES & \xmark & \cmark & \cmark & \xmark & \cmark & \cmark & \xmark & \cmark & \cmark & \xmark \\

Securify \cite{tsankov2018securify}& Fuzzer & YES & \xmark & \xmark & \xmark & \xmark & \xmark & \cmark & \xmark & \xmark & \cmark & \xmark \\

IR-Fuzz \cite{liu2023rethinking}& Fuzzer & YES & \xmark & \cmark & \xmark & \cmark & \cmark & \cmark & \xmark & \cmark & \xmark & \xmark \\

\textbf{LLAMA} & \textbf{Fuzzer} & \textbf{YES} & \cmark & \cmark & \cmark & \cmark & \cmark & \cmark & \cmark & \cmark & \cmark & \cmark \\
\hline
\end{tabular}\\
\footnotesize{AF: Assertion Failure; BD: Block State Dependency; EL: Ether Leak; FE: Freezing Ether; IO: Integer Overflows; RE: Reentrancy; TD: Transaction Order Dependency; UD: Unsafe Delegatecall; UE: Unhandled Exception; US: Unprotected Selfdestruct.} 
\label{Tab:comparison_vul}
\end{table*}

\subsection{Evaluation on Mutation Operators Selection}
We evaluated the effectiveness of different mutation operators in LLAMA. 
For each experiment, we applied a single mutation operator and recorded its impact on coverage growth as well as the number of times each operator was called.
As shown in Fig.~\ref{fig:execution-1}, the results demonstrated that different mutation operators had significantly different effects on coverage growth.
Some operators significantly increased coverage, while others had a much weaker effect, contributing little to the discovery of new execution paths.
Fig.~\ref{fig:execution-2} showed the number of times each mutation operator was called.
Resource wastes were found since some operators were repeatedly invoked during the fuzz testing process.
We observed that inefficient mutation operators failed to effectively improve coverage and consumed additional testing resources, thereby affecting the overall efficiency of fuzz testing.
The results suggested that inefficient mutation operators represent a bottleneck in current smart contract fuzz testing, wasting computational resources and reducing testing performance.
LLAMA addressed this issue by improving mutation scheduling through evolutionary strategies, leading to further performance enhancement.

\begin{table}[t]
\centering
\caption{Comparison of Mutation Types for SC Fuzzers}
\resizebox{0.5\textwidth}{!}{
\begin{tabular}{cccc|cccc} 
\hline
\textbf{Fuzzer}         & \textbf{TAM} & \textbf{EPM} & \textbf{TSM} & \textbf{Fuzzer}           & \textbf{TAM} & \textbf{EPM} & \textbf{TSM} \\ 
\hline
ContractFuzzer \cite{jiang2018contractfuzzer} & \xmark & \xmark & \xmark & Reguard \cite{liu2018reguard}        & \cmark & \xmark & \xmark \\
Echidna \cite{grieco2020echidna}        & \cmark & \xmark & \xmark & Harvey \cite{wustholz2020harvey}         & \cmark & \xmark & \xmark \\
IR-Fuzz \cite{liu2023rethinking}        & \cmark & \cmark & \cmark & sFuzz \cite{nguyen2020sfuzz}         & \cmark & \cmark & \xmark \\
ILF \cite{he2019learning}            & \cmark & \cmark & \xmark & Smartian \cite{choi2021smartian}      & \cmark & \cmark & \cmark \\
RLF \cite{su2022effectively}            & \cmark & \cmark & \xmark & SmartGift \cite{zhou2021smartgift}     & \xmark & \xmark & \xmark \\
xFuzz \cite{xue2022xfuzz}         & \cmark & \cmark & \xmark  &  \textbf{LLAMA} & \cmark & \cmark & \cmark\\
\hline
\end{tabular}
}
\footnotesize{Smart Contract: SC, Transaction Argument Mutation: TAM, Environmental Properties Mutation: EPM, and Transaction Sequence Mutation: TSM}
\label{Tab:comparison_mutation}
\end{table}

\subsection{Comparison on Tools of Vulnerability Detection}

We conducted a comparison of vulnerability types supported by various representative tools in the fields of static analysis, symbolic execution, and fuzz testing, as shown in Table~\ref{Tab:comparison_vul}.
The types and numbers of vulnerabilities covered by each tool varied. 
LLAMA led mainstream solutions in terms of supported vulnerability types, which covered ten types of vulnerabilities. 
It evidenced that LLAMA covered a broader range of vulnerability attack scenarios than most other tools.

Different mutation strategies were compared in the mutation process among various fuzzing methods, covering TAM, EPM, and TSM (refer to Table~\ref{Tab:comparison_mutation}).
TAM modified function arguments through techniques such as bit flipping, value incrementing, or random insertion \cite{wustholz2020harvey}.
EPM targeted blockchain-specific attributes to simulate different execution environments, such as timestamps, block numbers, or caller addresses \cite{liao2019soliaudit}.
TSM reordered transactions to trigger different state transitions \cite{olsthoorn2022syntest}.
The results showed that most fuzzing tools adopted TAM or EPM, while only a few incorporated TSM.
LLAMA supported all three mutation types and made it fully explore diverse execution paths and optimize the selection of mutation operators. 

%% file: Source/Coc.tex
\section{Conclusions} \label{sec: coc}

In this paper, we proposed a novel multi-feedback fuzzing framework for smart contracts, called LLAMA, which integrated LLM-guided initial seed generation with a feedback-driven evolutionary mutation strategy to enhance both coverage and vulnerability detection.
Our approach employed a hierarchical prompting mechanism to guide LLMs in generating semantically valid and vulnerability-sensitive transaction sequences.
A lightweight pre-fuzzing phase was proposed to select high-quality seeds to accelerate early-stage path exploration.
To improve efficiency, we incorporated a multi-feedback optimization strategy that adjusted operator probabilities based on historical feedback and uniform scheduling.



%% file: Main.bbl
\begin{thebibliography}{10}
\providecommand{\url}[1]{#1}
\csname url@samestyle\endcsname
\providecommand{\newblock}{\relax}
\providecommand{\bibinfo}[2]{#2}
\providecommand{\BIBentrySTDinterwordspacing}{\spaceskip=0pt\relax}
\providecommand{\BIBentryALTinterwordstretchfactor}{4}
\providecommand{\BIBentryALTinterwordspacing}{\spaceskip=\fontdimen2\font plus
\BIBentryALTinterwordstretchfactor\fontdimen3\font minus \fontdimen4\font\relax}
\providecommand{\BIBforeignlanguage}[2]{{%
\expandafter\ifx\csname l@#1\endcsname\relax
\typeout{** WARNING: IEEEtran.bst: No hyphenation pattern has been}%
\typeout{** loaded for the language `#1'. Using the pattern for}%
\typeout{** the default language instead.}%
\else
\language=\csname l@#1\endcsname
\fi
#2}}
\providecommand{\BIBdecl}{\relax}
\BIBdecl

\bibitem{zheng2020overview}
Z.~Zheng, S.~Xie, H.-N. Dai, W.~Chen, X.~Chen \emph{et~al.}, ``An overview on smart contracts: Challenges, advances and platforms,'' \emph{Future Generation Computer Systems}, vol. 105, pp. 475--491, 2020.

\bibitem{khan2021blockchain}
S.~N. Khan, F.~Loukil, C.~Ghedira-Guegan, E.~Benkhelifa, and A.~Bani-Hani, ``Blockchain smart contracts: Applications, challenges, and future trends,'' \emph{Peer-to-peer Netw. and App.}, vol.~14, pp. 2901--2925, 2021.

\bibitem{kannengiesser2021challenges}
N.~Kannengiesser, S.~Lins, C.~Sander, K.~Winter \emph{et~al.}, ``Challenges and common solutions in smart contract development,'' \emph{IEEE Trans. on Software Eng.}, vol.~48, no.~11, pp. 4291--4318, 2021.

\bibitem{chu2023survey}
H.~Chu, P.~Zhang, H.~Dong, Y.~Xiao \emph{et~al.}, ``A survey on smart contract vulnerabilities: Data sources, detection and repair,'' \emph{Info. \& Software Tech.}, vol. 159, p. 107221, 2023.

\bibitem{mehar2019understanding}
M.~I. Mehar, C.~L. Shier, A.~Giambattista, E.~Gong, G.~Fletcher \emph{et~al.}, ``Understanding a revolutionary and flawed grand experiment in blockchain: the dao attack,'' \emph{J. of Cases on Infor. Tech.}, vol.~21, no.~1, pp. 19--32, 2019.

\bibitem{vidal2024vulnerability}
F.~R. Vidal, N.~Ivaki, and N.~Laranjeiro, ``Vulnerability detection techniques for smart contracts: A systematic literature review,'' \emph{Journal of Systems and Software}, p. 112160, 2024.

\bibitem{li2018fuzzing}
J.~Li, B.~Zhao, and C.~Zhang, ``Fuzzing: a survey,'' \emph{Cybersecurity}, vol.~1, pp. 1--13, 2018.

\bibitem{sun2025adversarial}
J.~Sun, Z.~Yin, H.~Zhang, X.~Chen, and W.~Zheng, ``Adversarial generation method for smart contract fuzz testing seeds guided by chain-based llm,'' \emph{Auto. Software Eng.}, vol.~32, no.~1, pp. 1--28, 2025.

\bibitem{lemieux2023codamosa}
C.~Lemieux, J.~P. Inala, S.~K. Lahiri, and S.~Sen, ``Codamosa: Escaping coverage plateaus in test generation with pre-trained large language models,'' in \emph{IEEE/ACM 45th Int'l Conf. Software Eng.}\hskip 1em plus 0.5em minus 0.4em\relax IEEE, 2023, pp. 919--931.

\bibitem{ma2024combining}
W.~Ma, D.~Wu, Y.~Sun, T.~Wang, S.~Liu, J.~Zhang, Y.~Xue, and Y.~Liu, ``Combining fine-tuning and llm-based agents for intuitive smart contract auditing with justifications,'' \emph{arXiv preprint arXiv:2403.16073}, 2024.

\bibitem{meng2024large}
R.~Meng, M.~Mirchev, M.~B{\"o}hme, and A.~Roychoudhury, ``Large language model guided protocol fuzzing,'' in \emph{Proceedings of the 31st Ann'l Netw. and Distr. Sys. Security Symp.}, vol. 2024, 2024.

\bibitem{xia2024fuzz4all}
C.~S. Xia, M.~Paltenghi, J.~Le~Tian, M.~Pradel, and L.~Zhang, ``Fuzz4all: Universal fuzzing with large language models,'' in \emph{IEEE/ACM 46th Int'l Conf. on Software Eng.}, 2024, pp. 1--13.

\bibitem{qian2024mufuzz}
P.~Qian, H.~Wu, Z.~Du, T.~Vural, D.~Rong \emph{et~al.}, ``{MuFuzz:} sequence-aware mutation and seed mask guidance for blockchain smart contract fuzzing,'' in \emph{40th Int'l Conf. on Data Eng.}\hskip 1em plus 0.5em minus 0.4em\relax IEEE, 2024, pp. 1972--1985.

\bibitem{zhu2022fuzzing}
X.~Zhu, S.~Wen, S.~Camtepe, and Y.~Xiang, ``Fuzzing: a survey for roadmap,'' \emph{ACM Computing Surveys}, vol.~54, no. 11s, pp. 1--36, 2022.

\bibitem{wu2024we}
S.~Wu, Z.~Li, L.~Yan, W.~Chen, M.~Jiang \emph{et~al.}, ``Are we there yet? unraveling the state-of-the-art smart contract fuzzers,'' in \emph{Proceedings of the IEEE/ACM 46th Int'l Conf. on Software Eng.}, 2024, pp. 1--13.

\bibitem{herrera2021seed}
A.~Herrera, H.~Gunadi, S.~Magrath, M.~Norrish \emph{et~al.}, ``Seed selection for successful fuzzing,'' in \emph{30th ACM SIGSOFT Int'l Symp. Software Testing \& Analysis}, 2021, pp. 230--243.

\bibitem{andesta2020testing}
E.~Andesta, F.~Faghih, and M.~Fooladgar, ``Testing smart contracts gets smarter,'' in \emph{10th Int'l Conf. Comp. Know. Eng.}\hskip 1em plus 0.5em minus 0.4em\relax IEEE, 2020, pp. 405--412.

\bibitem{nguyen2020sfuzz}
T.~D. Nguyen, L.~H. Pham, J.~Sun, Y.~Lin, and Q.~T. Minh, ``sfuzz: An efficient adaptive fuzzer for solidity smart contracts,'' in \emph{ACM/IEEE 42nd Int'l Conf. on Software Eng.}, 2020, pp. 778--788.

\bibitem{choi2021smartian}
J.~Choi, D.~Kim, S.~Kim, G.~Grieco \emph{et~al.}, ``Smartian: Enhancing smart contract fuzzing with static and dynamic data-flow analyses,'' in \emph{36th IEEE/ACM Int'l Conf. Auto. Software Eng.}\hskip 1em plus 0.5em minus 0.4em\relax IEEE, 2021, pp. 227--239.

\bibitem{ji2021increasing}
S.~Ji, J.~Dong, J.~Qiu, B.~Gu \emph{et~al.}, ``Increasing fuzz testing coverage for smart contracts with dynamic taint analysis,'' in \emph{21st Int'l Conf. Software Quality, Reliability and Security}.\hskip 1em plus 0.5em minus 0.4em\relax IEEE, 2021, pp. 243--247.

\bibitem{tsankov2018securify}
P.~Tsankov, A.~Dan, D.~Drachsler-Cohen, A.~Gervais \emph{et~al.}, ``Securify: Practical security analysis of smart contracts,'' in \emph{ACM SIGSAC Conf. Comp. \& Comm. Sec.}, 2018, pp. 67--82.

\bibitem{tikhomirov2018smartcheck}
S.~Tikhomirov, E.~Voskresenskaya, I.~Ivanitskiy, R.~Takhaviev \emph{et~al.}, ``Smartcheck: Static analysis of ethereum smart contracts,'' in \emph{1st Int'l WKSP Emerging Trends Software Eng. for Blockchain}, 2018, pp. 9--16.

\bibitem{wang2020contractward}
W.~Wang, J.~Song, G.~Xu, Y.~Li \emph{et~al.}, ``Contractward: Automated vulnerability detection models for ethereum smart contracts,'' \emph{IEEE Trans. Netw. Sci. \& Eng.}, vol.~8, no.~2, pp. 1133--1144, 2020.

\bibitem{mossberg2019manticore}
M.~Mossberg, F.~Manzano, E.~Hennenfent, A.~Groce \emph{et~al.}, ``Manticore: A user-friendly symbolic execution framework for binaries and smart contracts,'' in \emph{34th IEEE/ACM Int'l Conf. Auto. Software Eng.}\hskip 1em plus 0.5em minus 0.4em\relax IEEE, 2019, pp. 1186--1189.

\bibitem{kalra2018zeus}
S.~Kalra, S.~Goel, M.~Dhawan, and S.~Sharma, ``Zeus: analyzing safety of smart contracts.'' in \emph{NDSS}, 2018, pp. 1--12.

\bibitem{feist2019slither}
J.~Feist, G.~Grieco, and A.~Groce, ``Slither: a static analysis framework for smart contracts,'' in \emph{IEEE/ACM 2nd Int'l WKSP Emerging Trends in Software Eng. for Blockchain}.\hskip 1em plus 0.5em minus 0.4em\relax IEEE, 2019, pp. 8--15.

\bibitem{shou2023ityfuzz}
C.~Shou, S.~Tan, and K.~Sen, ``Ityfuzz: Snapshot-based fuzzer for smart contract,'' in \emph{32nd ACM SIGSOFT Int'l Symp. Software Testing \& Analysis}, 2023, pp. 322--333.

\bibitem{liu2018reguard}
C.~Liu, H.~Liu, Z.~Cao, Z.~Chen \emph{et~al.}, ``Reguard: finding reentrancy bugs in smart contracts,'' in \emph{40th Int'l Conf. Software Eng.: Companion Proceeedings}, 2018, pp. 65--68.

\bibitem{xue2022xfuzz}
Y.~Xue, J.~Ye, W.~Zhang, J.~Sun \emph{et~al.}, ``xfuzz: Machine learning guided cross-contract fuzzing,'' \emph{IEEE Trans. Dependable \& Sec. Comp.}, vol.~21, no.~2, pp. 515--529, 2022.

\bibitem{vacca2021systematic}
A.~Vacca, A.~Di~Sorbo, C.~A. Visaggio, and G.~Canfora, ``A systematic literature review of blockchain and smart contract development: Techniques, tools, and open challenges,'' \emph{J. Sys. \& Software}, vol. 174, p. 110891, 2021.

\bibitem{he2019learning}
J.~He, M.~Balunovi{\'c}, N.~Ambroladze, P.~Tsankov, and M.~Vechev, ``Learning to fuzz from symbolic execution with application to smart contracts,'' in \emph{ACM SIGSAC Conf. Comp. Comm. Sec.}, 2019, pp. 531--548.

\bibitem{torres2021confuzzius}
C.~F. Torres, A.~K. Iannillo, A.~Gervais, and R.~State, ``Confuzzius: A data dependency-aware hybrid fuzzer for smart contracts,'' in \emph{IEEE Euro. Symp. Sec. \& Privacy}.\hskip 1em plus 0.5em minus 0.4em\relax IEEE, 2021, pp. 103--119.

\bibitem{su2022effectively}
J.~Su, H.-N. Dai, L.~Zhao, Z.~Zheng, and X.~Luo, ``Effectively generating vulnerable transaction sequences in smart contracts with reinforcement learning-guided fuzzing,'' in \emph{37th IEEE/ACM Int'l Conf. Auto. Software Eng.}, 2022, pp. 1--12.

\bibitem{so2020verismart}
S.~So, M.~Lee, J.~Park, H.~Lee, and H.~Oh, ``Verismart: A highly precise safety verifier for ethereum smart contracts,'' in \emph{IEEE Symp. Sec. \& Privacy}.\hskip 1em plus 0.5em minus 0.4em\relax IEEE, 2020, pp. 1678--1694.

\bibitem{zhuang2021smart}
Y.~Zhuang, Z.~Liu, P.~Qian, Q.~Liu, and ohters, ``Smart contract vulnerability detection using graph neural networks,'' in \emph{29th Int'l Joint Conf. Artificial Intell.}, 2021, pp. 3283--3290.

\bibitem{durieux2020empirical}
T.~Durieux, J.~F. Ferreira, R.~Abreu, and P.~Cruz, ``Empirical review of automated analysis tools on 47,587 ethereum smart contracts,'' in \emph{ACM/IEEE 42nd Int'l Conf. Software Eng.}, 2020, pp. 530--541.

\bibitem{swcregistry}
SmartContractSecurity, ``Swc registry,'' Available at \url{https://swcregistry.io}, 2020, accessed: 2024-03-17.

\bibitem{nikolic2018finding}
I.~Nikoli{\'c}, A.~Kolluri, I.~Sergey, P.~Saxena, and A.~Hobor, ``Finding the greedy, prodigal, and suicidal contracts at scale,'' in \emph{34th Ann. Comp. Sec. App. Conf.}, 2018, pp. 653--663.

\bibitem{chen2021defectchecker}
J.~Chen, X.~Xia, D.~Lo, J.~Grundy, X.~Luo, and T.~Chen, ``Defectchecker: Automated smart contract defect detection by analyzing evm bytecode,'' \emph{IEEE Trans. Software Eng.}, vol.~48, no.~7, pp. 2189--2207, 2021.

\bibitem{mueller2017mythril}
B.~Mueller, ``Mythril: A security analysis tool for evm bytecode,'' \url{https://github.com/ConsenSys/mythril}, 2017, accessed: 2024-10-26.

\bibitem{torres2018osiris}
C.~F. Torres, J.~Sch{\"u}tte, and R.~State, ``Osiris: Hunting for integer bugs in ethereum smart contracts,'' in \emph{34th Ann. Comp. Sec. App. Conf.}, 2018, pp. 664--676.

\bibitem{luu2016making}
L.~Luu, D.-H. Chu, H.~Olickel, P.~Saxena, and A.~Hobor, ``Making smart contracts smarter,'' in \emph{ACM SIGSAC Conf. Comp. \& Comm. Security}, 2016, pp. 254--269.

\bibitem{krupp2018teether}
J.~Krupp and C.~Rossow, ``{teEther}: Gnawing at ethereum to automatically exploit smart contracts,'' in \emph{27th USENIX Sec. Symp.}, 2018, pp. 1317--1333.

\bibitem{jiang2018contractfuzzer}
B.~Jiang, Y.~Liu, and W.~K. Chan, ``Contractfuzzer: Fuzzing smart contracts for vulnerability detection,'' in \emph{33rd ACM/IEEE Int'l Conf. Auto. Software Eng.}, 2018, pp. 259--269.

\bibitem{wang2020oracle}
H.~Wang, Y.~Liu, Y.~Li, S.-W. Lin \emph{et~al.}, ``Oracle-supported dynamic exploit generation for smart contracts,'' \emph{IEEE Trans. Dependable \& Sec. Comp.}, vol.~19, no.~3, pp. 1795--1809, 2020.

\bibitem{grieco2020echidna}
G.~Grieco, W.~Song, A.~Cygan, J.~Feist, and A.~Groce, ``Echidna: effective, usable, and fast fuzzing for smart contracts,'' in \emph{29th ACM SIGSOFT Int'l Symp. Software Testing \& analysis}, 2020, pp. 557--560.

\bibitem{wustholz2020harvey}
V.~W{\"u}stholz and M.~Christakis, ``Harvey: A greybox fuzzer for smart contracts,'' in \emph{28th ACM Joint Meeting on Euro. Software Eng. Conf. \& Symp. Foundations of Software Eng.}, 2020, pp. 1398--1409.

\bibitem{liu2023rethinking}
Z.~Liu, P.~Qian, J.~Yang, L.~Liu \emph{et~al.}, ``Rethinking smart contract fuzzing: Fuzzing with invocation ordering and important branch revisiting,'' \emph{IEEE Trans. Info. Fore. \& Sec.}, vol.~18, pp. 1237--1251, 2023.

\bibitem{zhou2021smartgift}
T.~Zhou, K.~Liu, L.~Li, Z.~Liu \emph{et~al.}, ``Smartgift: Learning to generate practical inputs for testing smart contracts,'' in \emph{IEEE Int'l Conf. Software Maintenance and Evolution}.\hskip 1em plus 0.5em minus 0.4em\relax IEEE, 2021, pp. 23--34.

\bibitem{liao2019soliaudit}
J.-W. Liao, T.-T. Tsai, C.-K. He, and C.-W. Tien, ``Soliaudit: Smart contract vulnerability assessment based on machine learning and fuzz testing,'' in \emph{6th Int'l Conf. IoT: Sys., Mngt. \& Sec.}\hskip 1em plus 0.5em minus 0.4em\relax IEEE, 2019, pp. 458--465.

\bibitem{olsthoorn2022syntest}
M.~Olsthoorn, D.~Stallenberg, A.~Van~Deursen, and A.~Panichella, ``Syntest-solidity: Automated test case generation and fuzzing for smart contracts,'' in \emph{ACM/IEEE 44th Int'l Conf. Software Eng.: Companion Proc.}, 2022, pp. 202--206.

\end{thebibliography}
